\documentclass[12pt,aps,amsmath,latexsym,amsfonts]{article}
\usepackage{jheppub}

\usepackage{graphicx}
\usepackage{amsfonts,amssymb,amsmath}
\usepackage[hang]{subfigure}
\usepackage{epstopdf}

\usepackage{color}

\newcommand{\be}{\begin{equation}}
\newcommand{\ee}{\end{equation}}
\newcommand{\ba}{\begin{eqnarray}}
\newcommand{\ea}{\end{eqnarray}}

\newcommand{\n}[1]{\label{#1}}

\def\cA{{\cal \omega}}
\def\ft#1#2{{\textstyle{\frac{\scriptstyle #1}{\scriptstyle #2} } }}
\def\fft#1#2{{\frac{#1}{#2}}}

\title{Thermodynamics of horizons: de Sitter black holes and reentrant phase transitions}
\author[a,b]{David Kubiz\v n\'ak,}
\author[b]{Fil Simovic}
\affiliation[a]{Perimeter Institute for Theoretical Physics,\\
31 Caroline St. N.,Waterloo, Ontario N2L 2Y5, Canada}
\affiliation[b]{Department of Physics and Astronomy, University of Waterloo,\\
Waterloo, Ontario N2L 3G1, Canada}
\emailAdd{dkubiznak@perimeterinstitute.ca}
\emailAdd{fil.simovic@gmail.com}
\abstract{
In this paper we propose a straightforward method for understanding the thermodynamics of black holes in de Sitter space, one that
will allow us to study these black holes in a way that is analogous to the anti-de Sitter case. As per usual, we formulate
separate thermodynamic first laws for each horizon present in the spacetime, and study their thermodynamics as if
they were independent systems characterized by their own temperature.
That these systems are not entirely independent and various thermodynamic quantities in them are in fact `correlated' is reflected by the fact that their thermodynamics can be captured by a single Gibbs free energy-like thermodynamic potential. This quantity contains information
about possible phase transitions in the system and allows us to uncover a rich phase structure for de Sitter black holes. In particular, we discover reentrant phase transitions for Kerr-dS black holes in six dimensions, a phenomenon recently observed for their six dimensional AdS cousins.
}
\begin{document}
\maketitle

\section{Introduction}
 Since the early 1970's it has been known that the characteristic properties of black holes such as the surface gravity, horizon area, mass, etc. are related by a law analogous to the first law of thermodynamics \cite{bekenstein1973black, Bardeen:1973gs}. Ever since then the subject of {\em black hole thermodynamics} has been an active area of research that has been both exciting and insightful.
 In particular, significant attention has been devoted to the study of asymptotically {\em anti-de Sitter} (AdS) black holes. For these black holes, thermodynamic equilibrium is straightforward to define and a wealth of interesting phenomena emerge.  For example, a thermal radiation/large black hole first order phase transition was found by Hawking and Page
 for the Schwarzschild-AdS black hole  \cite{hawking1983thermodynamics} and
the gravitational analogue of the liquid/gas phase transition was observed in the phase diagram of the charged AdS black hole \cite{Chamblin:1999tk, Chamblin:1999hg, Cvetic:1999ne, Kubiznak:2012wp}. More exotic AdS black hole spacetimes may even demonstrate reentrant phase transitions \cite{Altamirano:2013ane} or admit a tricritical point in their phase diagram \cite{Altamirano:2013uqa} (see \cite{Altamirano:2014tva, Dolan:2014jva, Kubiznak:2016qmn} for a review of these developments). The study of phase transitions in AdS black hole spacetimes is especially interesting due to their dual conformal field theory (CFT) interpretation through the  AdS/CFT correspondence \cite{Witten:1998zw}.

More recently, a holographic duality between gravity in de Sitter space and conformal field theory has been proposed \cite{Strominger:2001pn}, spurring interest in the thermodynamics of  black holes in asymptotically {\em de Sitter} (dS) spacetimes. Asymptotically dS black holes are also of direct interest in cosmology.
Thermodynamics in dS space is, however, burdened with issues not present in the AdS case. The existence of a cosmological horizon in addition to the black hole horizon means that the system associated with an observer living between the horizons is in a non-equilibrium state---such an observer
%in between the horizons
would find herself in a thermodynamic system characterized by two temperatures.
%we have a thermodynamic system with two temperatures.
Additionally, the lack of a global timelike Killing vector outside the black hole prevents one from defining a good notion of the black hole mass
(see \cite{Ashtekar:2014zfa, Ashtekar:2015lla} for recent developments in this direction).

One way to deal with the thermodynamics of asymptotically dS black hole spacetimes is to formulate several separate thermodynamic first laws, one for {each ``physical'' horizon present in the spacetime} \cite{Cai:2001sn, Cai:2001tv, Sekiwa:2006qj, Dolan:2013ft} (see also \cite{Gomberoff:2003ea}).
{Specifically, let us consider a general rotating dS black hole with several $U(1)$ charges in $d$ dimensions.
Typically, such a black hole admits three horizons, located at real positive radii $r$ which are determined from the horizon condition, say $f(r)=0$.
}
{
The {\em cosmological horizon} (we use subscript $c$) is located at the largest positive root $r_c$,
the {\em black hole horizon} (denoted by subscript $b$) corresponds to the second largest positive root $r_b$, and the {\em inner horizon} (with subscript $i$) corresponds to the third largest positive $r_i$ (if it exists).
The first laws for the three horizons then read \cite{Dolan:2013ft}:\footnote{
The thermodynamics of the inner black hole horizon might seem rather formal. However, only after it is included in the thermodynamic description, thermodynamic quantities demonstrate a certain type of `universality', which is speculated to provide some
information about the microscopics of black holes, see e.g. \cite{Cvetic:2010mn, Castro:2012av, Page:2015gia}.
}
\ba
\delta M&=&T_{b} \delta S_{b}+\sum_k (\Omega^k_{b}-\Omega^k_{\infty})\delta J^k+ \sum_j(\Phi_{b}^j-\Phi_{\infty}^j)\delta Q^j\, {+\, V_{b}\delta P\,,}\label{firstBHb}\\
\delta M&=&-T_{c} \delta S_{c}+\sum_k (\Omega^k_{c}-\Omega^k_{\infty})\delta J^k+ \sum_j(\Phi_{c}^j-\Phi_{\infty}^j)\delta Q^j
{\, +\, V_{c}\delta P}\,,\label{firstBHc}\\
\delta M&=&-T_{i} \delta S_{i}+\sum_k (\Omega^k_{i}-\Omega^k_{\infty})\delta J^k+ \sum_j(\Phi_{i}^j-\Phi_{\infty}^j)\delta Q^j
{\, +\, V_{i}\delta P}\,.\label{firstBHi}
\ea
Here, $M$ represents a quantity that {\em would be} the {\em ADM mass} in asymptotically AdS and flat cases and for this reason we simply refer
to it as ``mass''$\!.$\footnote{In the dS case such a quantity is ``conserved in space'' (rather than in time) due to the spacelike character of the Killing field $\partial_t$ in the region near infinity \cite{Ghezelbash:2001vs}.}
The horizon temperatures $T_b$, $T_c$ and $T_i$ are all defined to be positive. $S_b$, $S_c$ and $S_i$ denote the horizon entropies, $\Omega$'s and $J$'s denote the angular velocities and momenta respectively, $\Phi$'s and $Q$'s stand for the electric potentials and charges, and the pressure $P$ is related to the positive cosmological constant $\Lambda$ according to %(denoting $d$ the total number of spacetime dimensions)
\be\label{P}
P=-\frac{\Lambda}{8\pi}=-\frac{(d-1)(d-2)}{16\pi l^2}<0\,,
\ee
as it would be for a perfect fluid stress-energy tensor. Treating the cosmological constant as a thermodynamic pressure in this way is an idea that has recently received much attention \cite{Kubiznak:2016qmn}. $V_c$, $V_b$ and $V_i$ are the
thermodynamic volumes---the quantities thermodynamically conjugate to $P$---{given by}
\be
{
V_c=\Bigl(\frac{\partial M}{\partial P}\Bigr)_{S_c, J^1,Q^1\dots }\,,\quad
V_b=\Bigl(\frac{\partial M}{\partial P}\Bigr)_{S_b, J^1,Q^1\dots }\,,\quad
V_i=\Bigl(\frac{\partial M}{\partial P}\Bigr)_{S_i, J^1,Q^1\dots }\,.
}
\ee

Starting from \eqref{firstBHb}--\eqref{firstBHi}, it is possible to formulate the so called `subtracted' first laws. For example, for an observer in between the cosmological and
the black hole horizon, the difference of \eqref{firstBHb} and \eqref{firstBHc} gives
\be
0=T_{b} \delta S_{b}+T_{c}\delta S_{c} +\sum_i (\Omega^i_{b}-\Omega^i_{c})\delta J^i
+\sum_j(\Phi_{b}^j-\Phi_c^j)\delta Q^j
-V \delta P\,,\label{first2}
\ee
 where $V$ stands for the subtracted volume of the `observable universe', $V=V_c-V_b\geq 0$. Although such subtracted laws provide useful information, in this paper we directly work with relations \eqref{firstBHb}--\eqref{firstBHi}.

The above three laws are accompanied by the corresponding {Smarr--Gibbs--Duhem} formulae
\ba
\frac{d-3}{d-2}M&=&T_{b}S_{b}+\frac{d-3}{d-2}\sum_j(\Phi_{b}^j-\Phi_{\infty}^j)Q^j
+\sum_{k}(\Omega_{b}^k-\Omega_{\infty}^k)J^k-\frac{2}{d-2}V_{b}P\,,\quad \label{Smarr1}\\
\frac{d-3}{d-2}M&=&-T_{c}S_{c}+\frac{d-3}{d-2}\sum_j(\Phi_{c}^j-\Phi_{\infty}^j)Q^j +\sum_{k}(\Omega_{c}^k-\Omega_{\infty}^k)J^k-\frac{2}{d-2}V_{c}P\,,\quad \label{Smarr2}\\
\frac{d-3}{d-2}M&=&-T_{i}S_{i}+\frac{d-3}{d-2}\sum_j(\Phi_{i}^j-\Phi_{\infty}^j)Q^j +\sum_{k}(\Omega_{i}^k-\Omega_{\infty}^k)J^k-\frac{2}{d-2}V_{i}P\,,\quad \label{Smarr2}
\ea
{which can be derived from the corresponding first laws \eqref{firstBHb}--\eqref{firstBHi}}
{via the dimensional scaling argument, see e.g. \cite{Kastor:2009wy}.}

Note that the three first laws \eqref{firstBHb}--\eqref{firstBHi} are not truly independent and the various thermodynamic quantities in them
are in fact `correlated' and %more importantly
determined from the same number of independent parameters as in the AdS case.
For example, consider the Schwarzschild-dS black hole spacetime in four dimensions, with the line element given by
\be
ds^2=-fdt^2+\dfrac{dr^2}{f}+r^2d\Omega_2^2\,,\quad f=1-\dfrac{2M}{r}+\frac{8\pi P r^2}{3}\,,
\ee
and $d\Omega^2_2$ denoting the element on the unit 2-sphere. By fixing the cosmological constant $P$ and the mass $M$, the
cosmological and black hole horizon radii $r_k$ $(k=b,c)$ are given by the largest ($r_c$) and second largest ($r_b$) positive roots of $f(r_k)=0$.
(Note that there is no inner horizon in this case.) Having fixed these radii, both temperatures are then uniquely determined by $T_k=|f'(r_k)|/(4\pi)$, as are the entropies $S_k=\pi r_k^2$, and thermodynamic volumes $V_k=\frac{4}{3}\pi r_k^3$. In other words, the complete thermodynamic description of all horizons is fixed by choosing $M$ and $P$, exactly as in the Schwarzschild-AdS case. This immediately generalizes to arbitrary dS black holes. {As we shall, see this will allow us to construct
a Gibbs free energy-like quantity that captures behavior of all three dS horizons and is in some sense a continuation of the AdS Gibbs to negative pressures and temperatures.
}
%It is because of this fact we believe a %n effective
%thermodynamic description analogous to the AdS case is possible for dS black holes.

The aim of this paper is to study the
thermodynamic behavior of asymptotically dS black holes and, in particular, to
investigate if such black holes admit phase transitions similar to their AdS cousins \cite{
hawking1983thermodynamics, Chamblin:1999tk, Chamblin:1999hg, Cvetic:1999ne, Kubiznak:2012wp, Altamirano:2013ane, Altamirano:2013uqa, Altamirano:2014tva, Dolan:2014jva, Kubiznak:2016qmn}.
%For concreteness and technical simplicity, in this paper we limit ourselves to illustrating these proposals for the four-dimensional Schwarzschild-dS and Reissner--Nordstrom-dS ``test geometries'' (reviewed in Sec.~2) and leave more complicated spacetimes to a future study \cite{Fil}.

One approach to such a task is that of the {\em effective temperature} \cite{Urano:2009xn, Ma:2013aqa, Zhao:2014zea, Zhao:2014raa, Ma:2014hna, Guo:2015waa}.\footnote{See also \cite{McInerney:2015xwa} and \cite{Li:2016zca} for recent alternative approaches.}
 In this approach one concentrates on  an observer who is located in an `observable part of the universe', in between the black hole horizon and the cosmological horizon. For such an observer a new effective first law is imposed such that the system is assigned a {\em `total entropy'} that equals the sum of the black hole horizon and the cosmological horizon entropies, see e.g. \cite{Kastor:1992nn, Bhattacharya:2015mja},
\be\label{Stotal}
S=S_b+S_c\,,
\ee
while the parameter $M$ is interpreted as internal energy, and the parameter $V=V_c-V_b$ as the observable volume.
This then defines the effective temperature and pressure---quantities thermodynamically conjugate to $S$ and $V$.
Interestingly, this prescription leads to a Schwarzschild-dS black hole analogue \cite{FilJa} of the famous Hawking--Page transition \cite{hawking1983thermodynamics}. However, in general the physical meaning or the positive-definiteness of the effective temperature and pressure are not obvious. Moreover, applying the same prescription to more complicated dS black hole spacetimes results in complicated phase diagrams whose interpretation is not clear \cite{FilJa}. %For this reason here we do not follow the effective temperature approach.

In this paper we propose an alternative path towards the thermodynamics of dS black holes. Namely,
we do not attempt to assign the system a single effective temperature. % but rather take seriously the thermodynamic behavior of each horizon.
Instead, we follow the following steps.
\begin{enumerate}
\item
We study the thermodynamics of all three physical horizons---treating them as {\em `independent thermodynamic systems'} in equilibrium, characterized by their own temperature, thermodynamic behavior, and first law \eqref{firstBHb}--\eqref{firstBHi}.\footnote{Let us stress that we do not claim that the entire dS black hole system is in thermodynamic equilibrium. That this is not the case is, for example, reflected by the fact that the corresponding Euclidean instanton is not regular unless the ratio of horizon temperatures is a rational number. Since each horizon is in general at a different temperature, there will be a heat flow in the system that might eventually, if the system were not disturbed, equilibrate the horizon temperatures. However, in what follows we assume that such a process would happen at a time scale which is much longer (due to the smallness of $\hbar$) than a typical timescale of thermodynamic phase transitions induced by an `external force' considered below.}

\item
Since the thermodynamic quantities on these horizons are (at least for a given classical solution) correlated, one can assign to the whole system a `single' Gibbs free energy-like quantity $G$.   This quantity, which is in some sense a continuation of the Gibbs free energy of AdS black holes to negative pressures and negative temperatures, captures the behavior of all three horizons.\footnote{The definition of $G$ is not an additional assumption but rather a mathematical convenience that directly follows from 1) and the correlation of thermodynamic quantities at different horizons. In our study we discard the possibility of thermal phase transitions governed by Euclidean instantons that would `change the character' of the classical gravitational solution under consideration, e.g. \cite{ginsparg1983semiclassical}.
}

\item
We assume that we have a mechanism (external force) at our disposal that allows us to tune thermodynamic parameters of the horizons.\footnote{This is a standard assumption made in black hole thermodynamics in order to describe possible phase transitions.
For example, in the famous Hawking--Page transition \cite{hawking1983thermodynamics} it is assumed that an infinite reservoir can `pump' energy into the system, changing radiation to a large black hole phase as the temperature of the system increases. A novel feature that seems to emerge in dS black hole spacetimes is that a change in thermodynamic parameters of one horizon necessary leads to a change of thermodynamic parameters of the other two horizons. Although strange at a first sight, exactly the same correlation {silently takes place in asymptotically flat or AdS black hole spacetimes} where the inner horizon is assumed to adjust accordingly to the change of the black hole horizon. For a discussion of this phenomenon for the charged AdS black holes see App.~\ref{appA}.
}

\item
Possible instabilities/phase transitions at each horizon, encoded in quantity $G$, are taken as a sign of instability/phase transition of the whole dS system.\footnote{That phase transitions can happen for systems with temperature gradients and far from thermal equilibrium is well documented by our everyday experience. Consider, for example, a typical ice fishing story: while we boil water to make tea and get warmer, our ice fishing hole freezes over.
}
\end{enumerate}
%Note that all the steps are in fact standardly used in black hole thermodynamics. For this reason we believe that this approach is more straightforward than for example that of the effective temperature.

Effectively, our procedure boils down to finding the Gibbs free energy $G$. % (extended to all black hole radii and hence negative temperatures).
%that captures the complete thermodynamic description of all three horizons (treated as independent thermodynamic systems). As we shall see this is in some sense a continuation of the Gibbs free energy of AdS black holes to negative pressures and negative temperatures.
By studying its properties one can understand the thermodynamic behavior of each horizon as if it were an independent thermodynamic system.
Putting all the information together, one can then infer possible black hole phase transitions.
%\footnote{The situation is very similar to the following picture. Imagine a thermodynamic system consisting of a room equipped with several heaters and inhabited by a tenant who has a mechanism to control one of the heaters, a `black hole horizon heater' for example. Such a tenant is obviously in a non-equilibrium state as the heat flows from one furnace to another and cannot use equilibrium thermodynamics nor define an effective temperature. However, during her experiments she observes an interesting fact. As she changes the temperature of the black hole horizon heater, other heaters, due to their secret hidden entanglement, also adjust their temperature accordingly. At some point it may happen that the state of the heaters is such that the system undergoes a phase transition (for example the air in the room suddenly condenses to a liquid drowning the poor obsrever). It is in this sense that we study thermodynamic phase transitions of dS black holes.}
Treated in this way we discover that dS black holes admit phase transitions and complicated phase diagrams, similar to the AdS case. For example, as we shall see, one finds a {\em reentrant phase transition} analogous to \cite{Altamirano:2013ane}.

The rest of the paper is organized as follows. In the next section we pedagogically illustrate the above procedure using the simple example of four-dimensional charged dS black holes, studying the thermodynamics of each horizon separately and showing how this can be described by a single thermodynamic function $G$. Sec.~3 is devoted to rotating dS black holes in four and higher dimensions; in particular the reentrant phase transitions are discovered for the 6-dimensional case within a certain range of parameters. Sec.~4 is devoted to conclusions. App.~A describes a similar construction applicable to AdS black holes. App.~B contains technical results regarding the rotating dS black holes in all dimensions.
%esctribe the

%%%%%%%%%%%%%%%%%%%%%%%%%%%%%%%%%%%%%%%%%%%%%%%%%%%%%%%%%%%%%%%%%%%%%%%%%%
%\section{Black hole horizon first law}
\section{Thermodynamics of three horizons: charged dS case}
In this section we %illustrate the above procedure on a few concrete examples. In particular,
show  how the thermodynamics of the three de Sitter horizons can be captured by a single Gibbs free energy, and how such a quantity encodes information about possible phase transitions of the de Sitter black hole system. For concreteness and technical simplicity this is illustrated on a `test spacetime' of the charged dS black hole in four dimensions.
% (reviewed in the next subsection). A more complicated example is briefly studied towards the end of this section to show that dS black holes can, similar to their AdS cousins, demonstrate reentrant phase transitions.

%%%%%%%%%%%%%%%%%%%%%%%%%%%%%%%%%%%%%%%%%%%%%%%%%%%%%%%%%%%%%%%%
\subsection{Charged dS black holes}

The charged dS black hole line element takes the following form:
\be\label{RNmetricdS}
ds^2=-fdt^2+\dfrac{dr^2}{f}+r^2d\Omega_2^2\,,\quad f=1-\dfrac{2M}{r}+\frac{Q^2}{r^2}+\frac{8\pi P r^2}{3}\,,
\ee
where $M$ stands for the mass and $Q$ for the electric charge. The Schwarzschild-dS metric is recovered by setting $Q=0$.
The cosmological $r_c$, black hole $r_b$, and inner $r_i$ (for $Q\neq 0$) horizons are located at $r_c\geq r_b\geq r_i\geq 0$, given by $f(r_k)=0$. The cases when two or more horizons coincide are special and are partly discussed below. The thermodynamic quantities associated with the three horizons are given by, see e.g. \cite{Dolan:2013ft},
\ba
T_b&=&\dfrac{1}{4\pi r_b}\Big(1+8\pi Pr_b^2-\dfrac{Q^2}{r_b^2}\Big)\,,\quad
S_b=\pi r_b^2\,,\quad \Phi_b=\dfrac{Q}{r_b}\,,\quad V_b=\dfrac{4}{3}\pi r_b^3\,,\nonumber\\
T_c&=&\dfrac{-1}{4\pi r_c}\Big(1+8\pi P r_c^2-\dfrac{Q^2}{r_c^2}\Big)\,,\quad
S_c=\pi r_c^2\,,\quad \Phi_c=\dfrac{Q}{r_c}\,,\quad V_c=\dfrac{4}{3}\pi r_c^3\,,\nonumber\\
T_i&=&\dfrac{-1}{4\pi r_i}\Big(1+8\pi P r_i^2-\dfrac{Q^2}{r_i^2}\Big)\,,\quad
S_i=\pi r_i^2\,,\quad \Phi_i=\dfrac{Q}{r_i}\,,\quad V_i=\dfrac{4}{3}\pi r_i^3\,,\label{quant}
\ea
and obey the three first laws \eqref{firstBHb}--\eqref{firstBHi}, as well as the corresponding Smarr relations.
%For the purpose of Sec.~4, we shall need
%the volume of the observable universe \eqref{Vtotal}, $V=V_c-V_b$, and its total entropy \eqref{Stotal}, $S=S_b+S_c$.

Several special cases of charged dS spacetimes are of importance. The {\em extremal black hole limit} corresponds to the case where the inner horizon coincides with the  black hole horizon, $r_b=r_i$. It can be easily shown that in this case we have $T_b=T_i=0$. The second important case is the {\em Nariai limit} in which $r_b=r_c$,  and consequently, $T_b=T_c=0$. Finally, in the limit $r_b\to 0$ we recover the pure dS case, while in the limit $r_c\to \infty$ the asymptotically flat black hole solution is obtained. We refer to \cite{nariai1951new, ginsparg1983semiclassical, Mann:1995vb, Booth:1998gf, Anninos:2010gh} for more details on these special cases.

\subsection{Black hole horizon}

Let us start our discussion by studying the  behavior of the {\em black hole horizon}.
The corresponding thermodynamic system is characterized by the black hole horizon temperature $T_b$, while the first law \eqref{firstBHb} now reads
\be
\delta M=T_{b} \delta S_{b}+\Phi_{b}\delta Q\, {+\, V_{b}\delta P\,.}
\ee
The structure of this formula suggests that the mass $M$ should be treated as the {\em gravitational enthalpy}, {$M=H_b$,} similar to the AdS case \cite{Kastor:2009wy}.
In fact, the above first law is exactly identical to the first law of AdS black holes, see e.g. \cite{Altamirano:2014tva}, with the exception that $P=-\Lambda/8\pi$ is now negative.
For this reason one might expect that the thermodynamic behavior of the dS black hole horizon can be obtained by simply `continuing' the AdS thermodynamics to negative pressures. The quantity analogous to the Gibbs free energy $G=G(T_b, P, Q)$ is given by the Legendre transformation of the enthalpy $M$,
\be\label{BHGibbs}
{G=H_b-T_bS_b=M-T_{b}S_b\,.}
\ee
Note that $T_b$ is manifestly non-negative and vanishes in both the Nariai limit and the extremal black hole limits.

\begin{figure}
\centering
\begin{tabular}{cc}
{\includegraphics[width=0.47\textwidth,height=0.27\textheight]{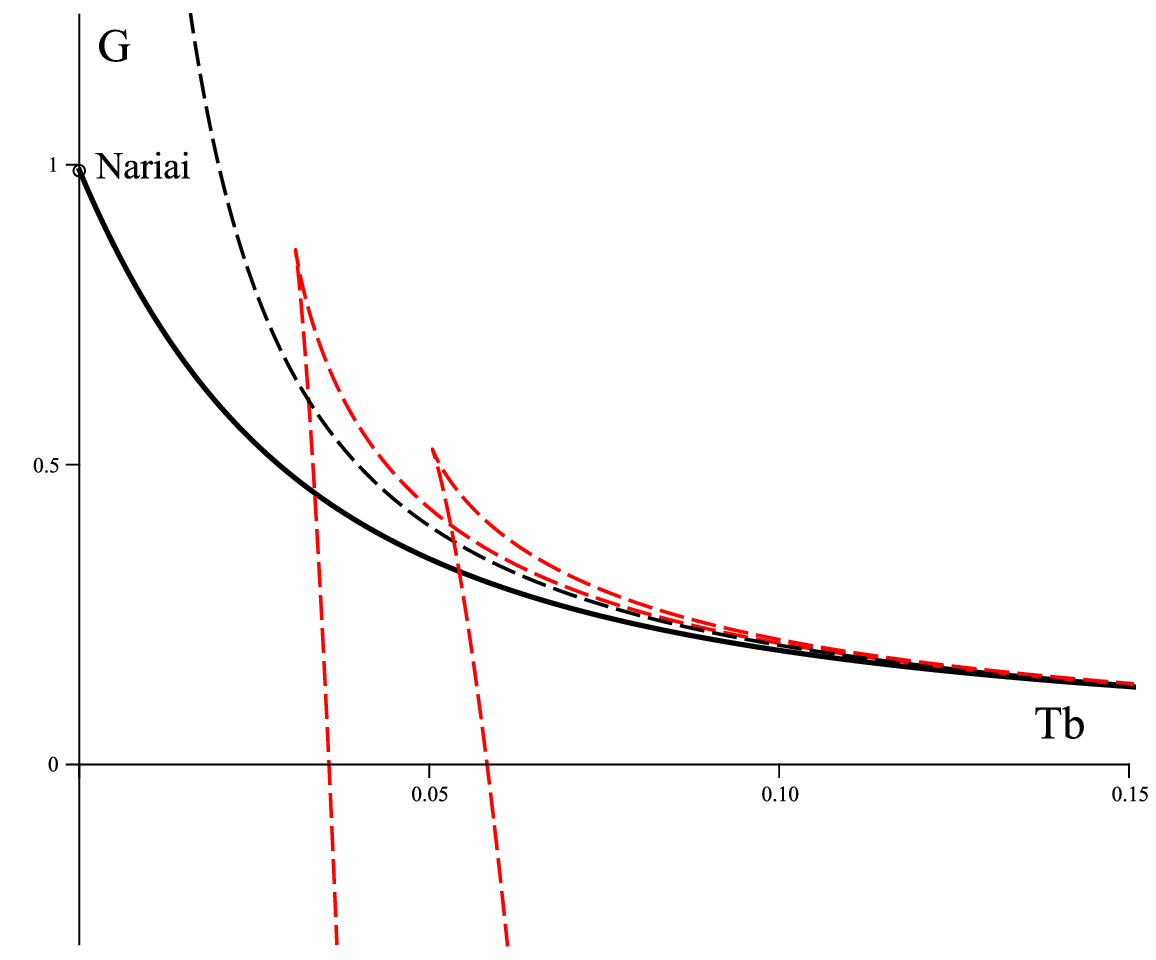}} &
%\rotatebox{-90}{
\includegraphics[width=0.47\textwidth,height=0.27\textheight]{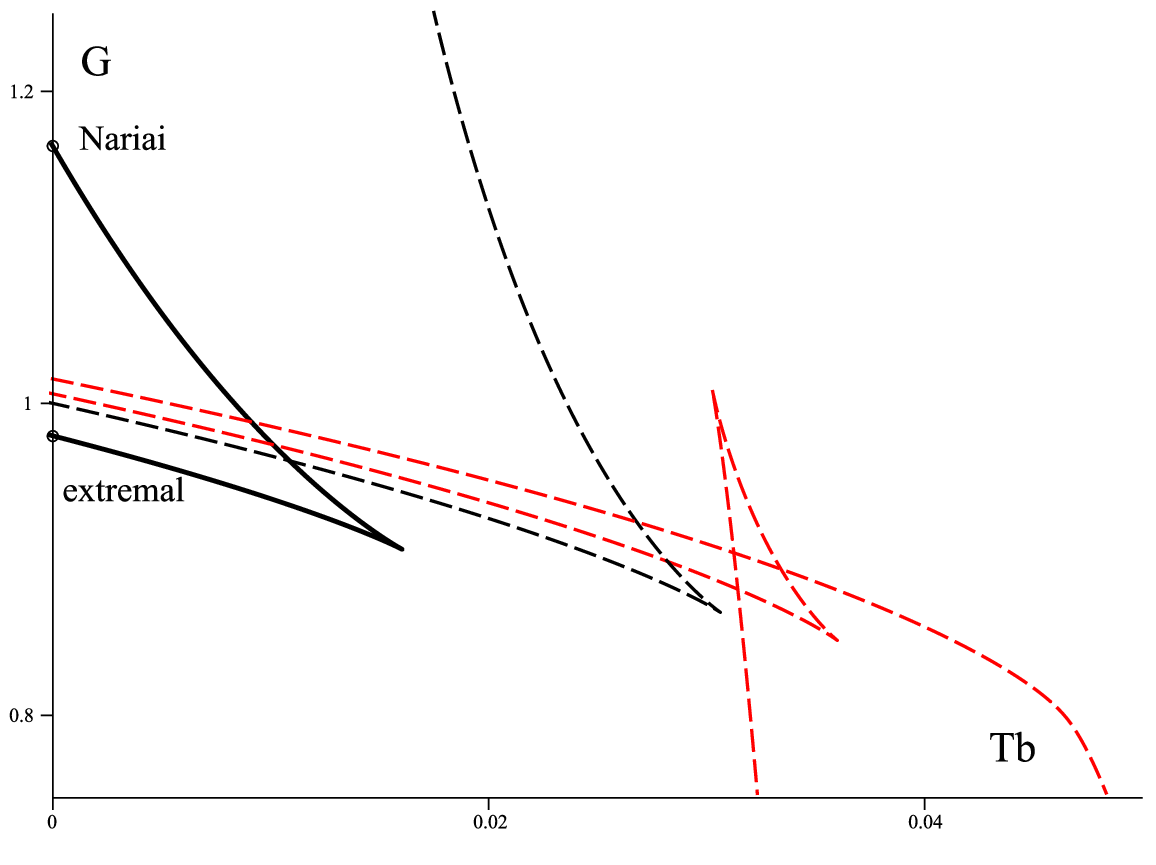}\\
\end{tabular}
\caption{{\bf Gibbs free energy: black hole horizon.}
The Schwarzschild (left) and the charged with $Q=1$ (right) cases are displayed for various pressures.
The thick black curve corresponds to the dS case with $P=-0.0045$, the dashed black curve to the asymptotically flat case with $P=0$, and
the red dashed lines to the AdS case with $P=0.0015$ (below $P_c$) and $P=0.0040$ (above $P_c$), respectively. Clearly the dS case can be understood as a continuation of the AdS case to the region of negative pressures. Since the critical point (in the charged case) occurs for positive pressures, there is no criticality in the dS case. Note however that the upper branch in between the Nariai limit and the cusp is thermodynamically unstable as it has negative specific heat and higher Gibbs free energy than the lower branch that terminates at the extremal limit.
}
\label{fig1}
\end{figure}
Having defined the Gibbs free energy, we can display its behavior as a function of the temperature $T_b$ and reveal the associated thermodynamic behavior. For our test spacetime this is displayed by thick black curves in Fig.~\ref{fig1}. In the charged case we observe two branches of black holes (meeting at a cusp), compared to one branch in the $Q=0$ limit. This behavior of $G$ can be understood as the continuation of the known behavior of the Gibbs free energy in the AdS case, displayed in the figure by dashed red curves, and, apart from the `Nariai cut-off', it is in many respects similar to the behavior of the Gibbs free energy of asymptotically flat spacetimes, c.f. dashed black curves.
Whereas the charged AdS case (see App.~\ref{appA} for more details) admits a critical point characterized by a critical pressure $P_c=1/96\pi Q^2$,
and below such pressure one observes a small black hole/large black hole phase transition, in the case of charged dS black holes no such behavior exists. There is no swallow tail and the behavior of $G$ is relatively simple: two branches (the upper one being thermodynamically unstable) meet at a cusp.

{
To probe the local thermodynamic stability, one can study the specific heat. For the black hole horizon this is naturally defined as
\be
C_b=\frac{\partial H_b}{\partial T_b}\Bigr|_{P,Q}=\frac{\partial M}{\partial T_b}\Bigr|_{P,Q}\,.
\ee
As expected, for the Schwarzschild-dS case $C_b$ is negative and vanishes in the Nariai and dS limits, see left Fig.~\ref{fig:CS}.}
{In the charged dS case, the black hole branch with lower Gibbs (in between the extremal black hole and the cusp) has positive $C_b$ and is locally thermodynamically stable, at the cusp $C_b$ suffers from an infinite jump, and is negative for the upper branch of black holes (in between the cusp and the Nariai limit). In the Nariai and extremal limits $C_b$ vanishes, see left Fig.~\ref{fig:CRN}. %Both charged and uncharged cases are reminiscent of what happens in the asymptotically flat case.
}

\begin{figure}
\centering
\begin{tabular}{cc}
{\includegraphics[width=0.47\textwidth,height=0.25\textheight]{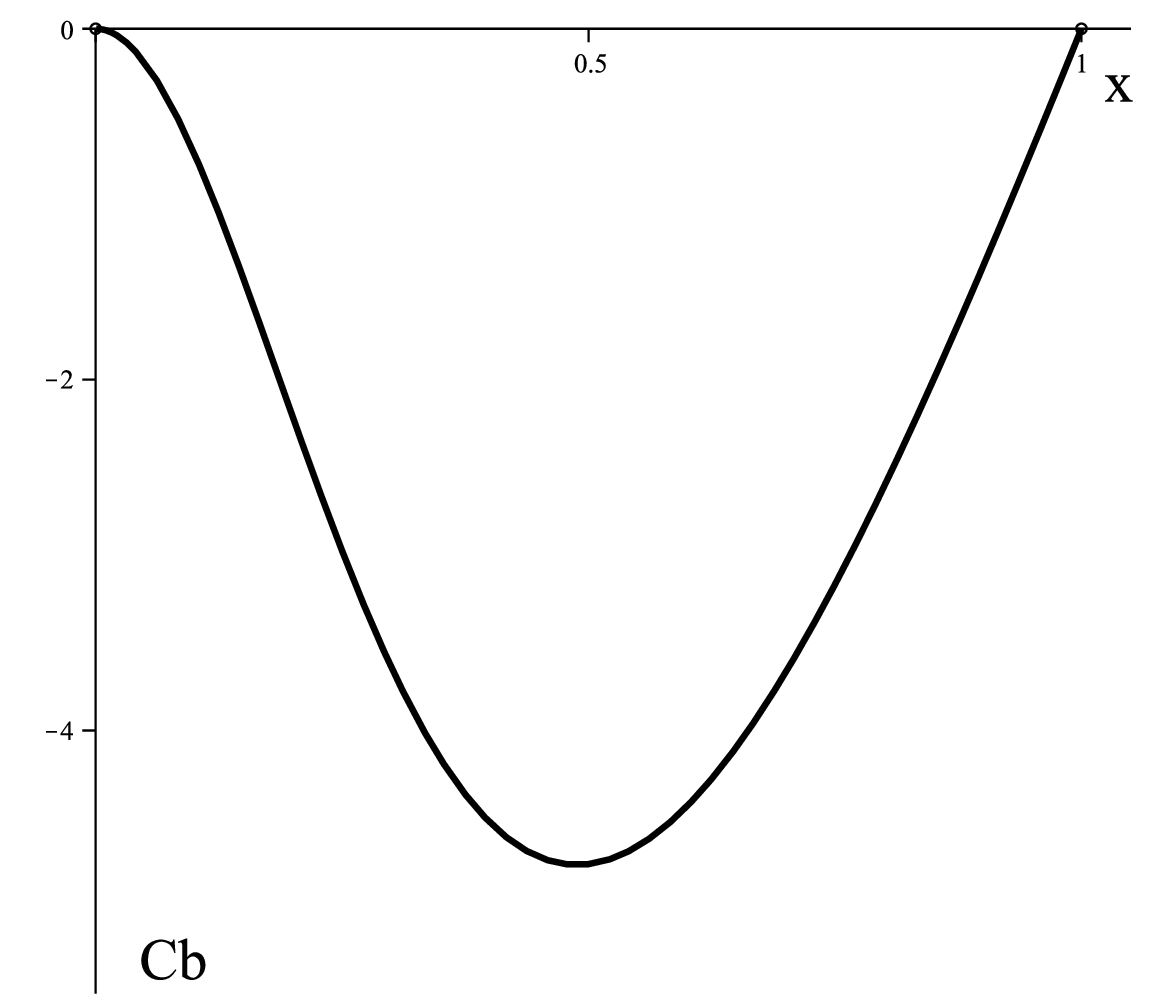}} &
%\rotatebox{-90}{
{\includegraphics[width=0.47\textwidth,height=0.25\textheight]{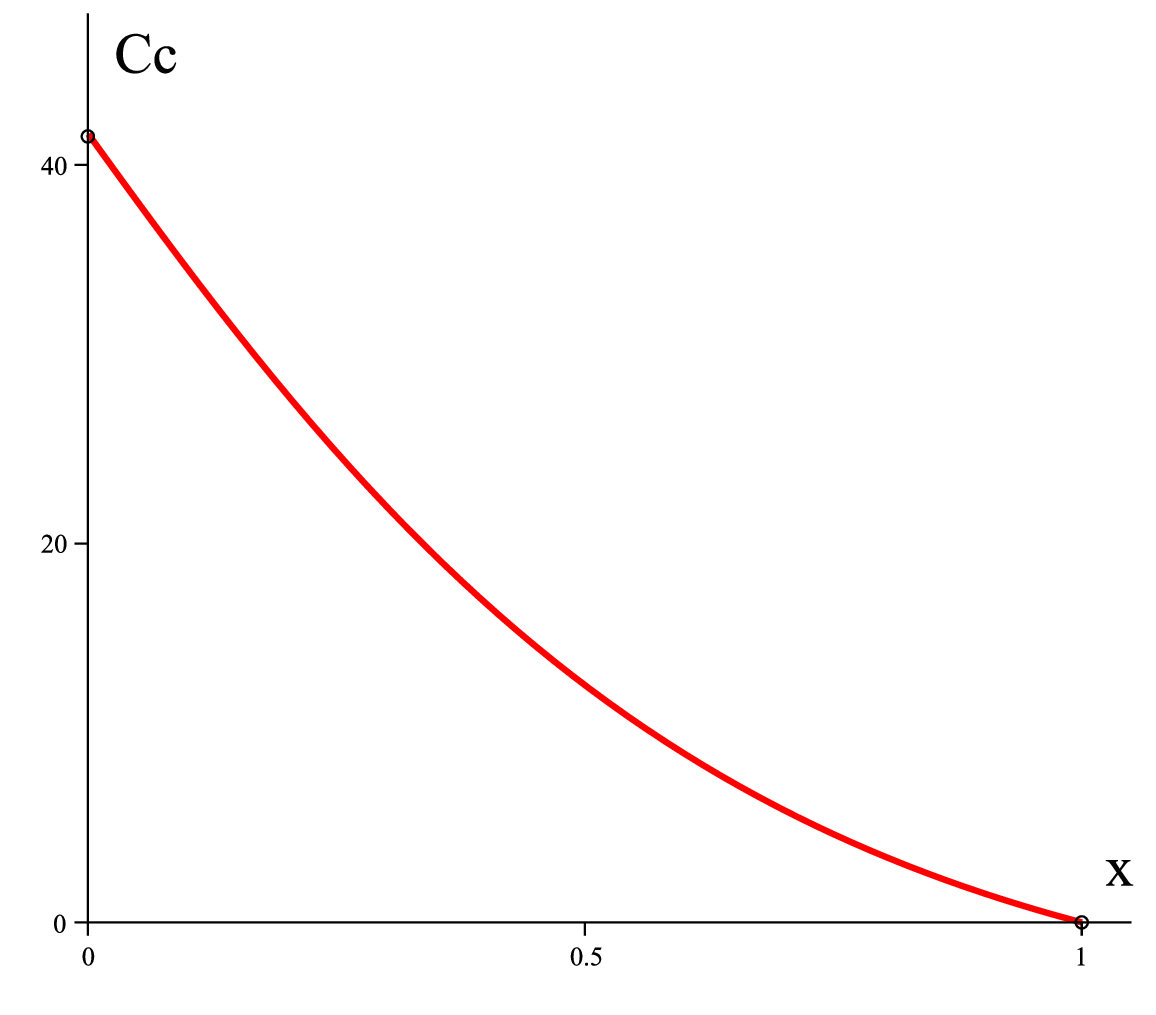}} \\
\end{tabular}
\caption{
{{\bf Specific heats: Schwarzschild-dS case.}
We display the two specific heats: $C_b$ (left) and $C_c$ (right) for $P=-0.009$ as a function of $x=r_b/r_c$.
$C_b$ is non-positive and vanishes in both the Nariai ($x=1)$ and dS ($x=0$) limits. $C_c$ is non-negative and vanishes in the Nariai limit. }
}
\label{fig:CS}
\end{figure}

\begin{figure}
\centering
\begin{tabular}{ccc}
{\includegraphics[width=0.32\textwidth,height=0.22\textheight]{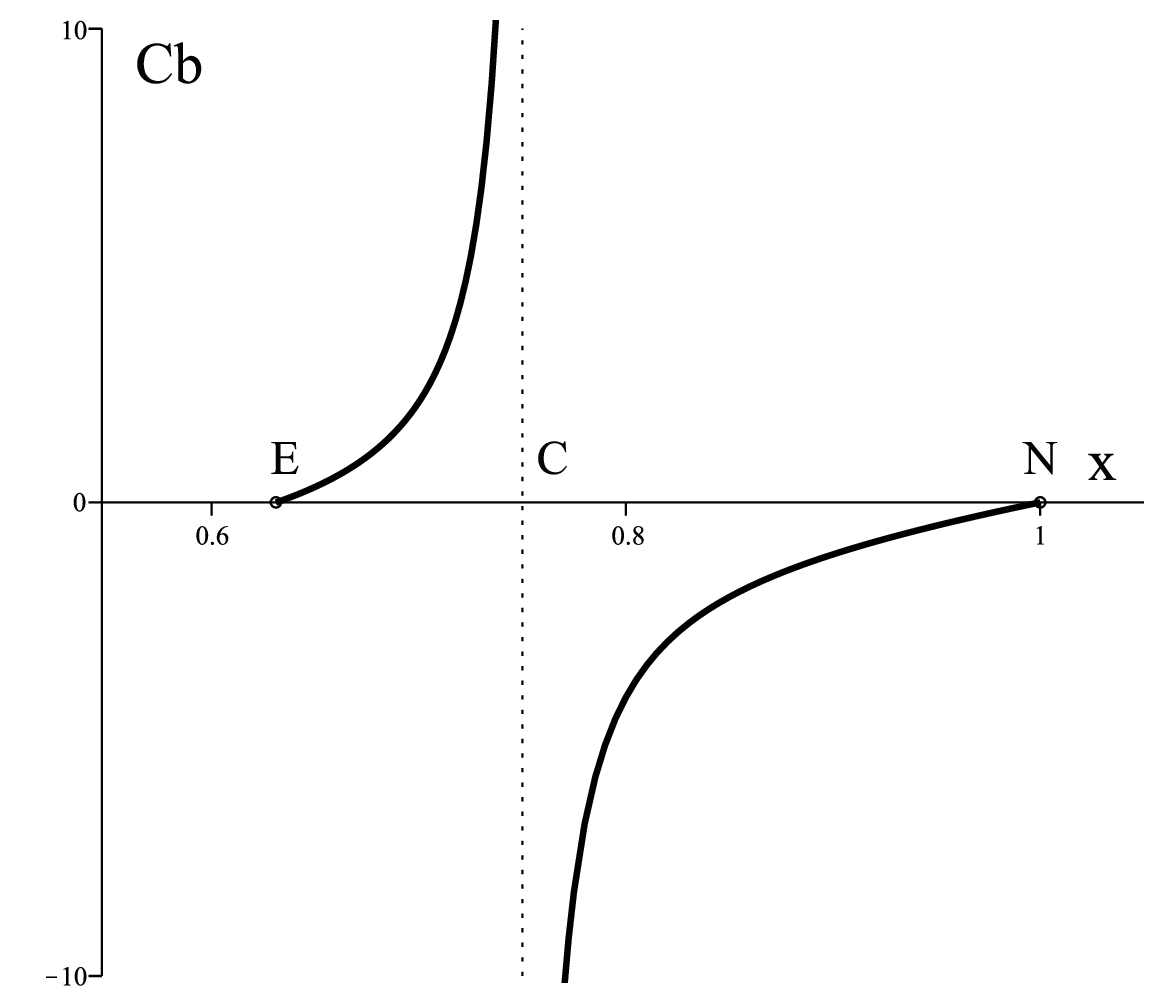}} &
%\rotatebox{-90}{
{\includegraphics[width=0.32\textwidth,height=0.22\textheight]{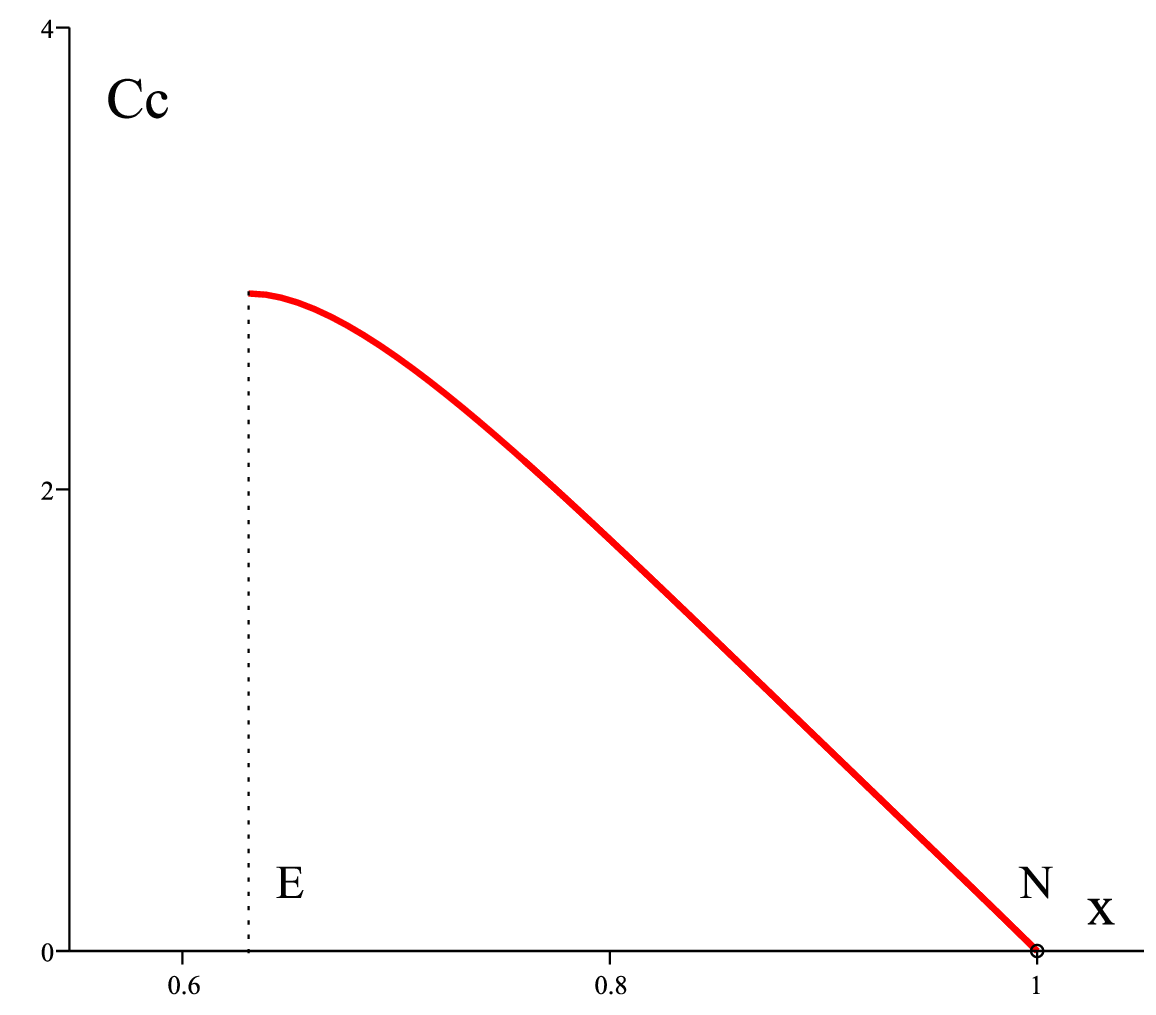}} &
\includegraphics[width=0.32\textwidth,height=0.22\textheight]{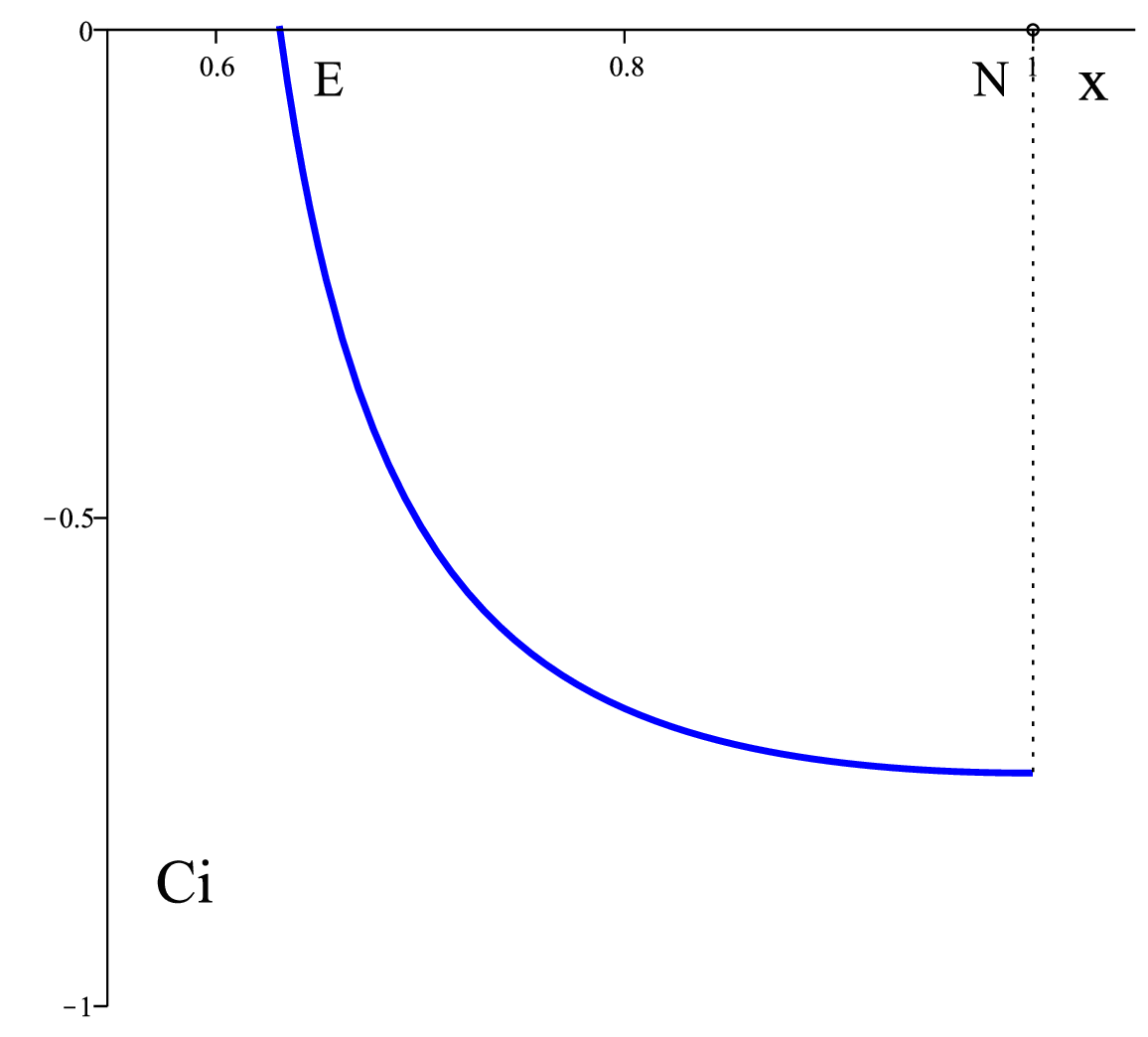}\\
\end{tabular}
\caption{{{\bf Specific heats: charged dS case.}
We display the three specific heats: $C_b$ (left), $C_c$ (middle) and $C_i$ (right) for $Q=1$ and
$P=-0.009$. $C_b$ suffers from an infinite jump at $x$ corresponding to the cusp (C) of the Gibbs free energy
and vanishes in both the Nariai (N) and extremal (E) limits.
$C_c$ vanishes at the Nariai limit and otherwise is manifestly positive, $C_i$ vanishes in the extremal limit and otherwise is negative.
}}
\label{fig:CRN}
\end{figure}

\subsection{Cosmological and inner horizons}
Let us next investigate the thermodynamic behavior of the {\em cosmological horizon}, characterized by the temperature $T_c$. Staring at
Eq.~\eqref{firstBHc}, we realize that the $T_c dS_c$ term has an opposite sign to what is customary in order to treat $M$ as the enthalpy.
To fix this, we set $H_c=-M<0$, and write (note the flipped sign of the work terms)
\be
\delta H_c=T_{c} \delta S_{c}- \Phi_{c}\delta {Q}-V_{c}\delta {P}\,.\label{firstBHd}
\ee
In other words, we identify the following Gibbs free energy:
\be
G_c=G_c(T_c, {P}, Q)=-M-T_{c}S_c\,.
\ee
For our test spacetimes, this is displayed in Fig.~\ref{fig1c}.
On its own, the behavior of $G_c$ seems `boring' and does not indicate the presence of any thermodynamically unstable regions or phase transitions. However, this is misleading as, similar to the AdS case discussed in App.~\ref{appA}, one has to `coordinate' the behavior of the cosmological horizon with the behavior of the black hole horizon, i.e., to consider $G$ and $G_c$ {\em simultaneously}. Consequently, for the charged dS black hole a part of the $G_c$ graph becomes thermodynamically unstable (displayed by dashed black curve), as it corresponds to the upper (thermodynamically unstable) branch of $G$, c.f. Figs.~\ref{fig1} and \ref{fig1c}.

\begin{figure}
\centering
\begin{tabular}{cc}
{\includegraphics[width=0.47\textwidth,height=0.27\textheight]{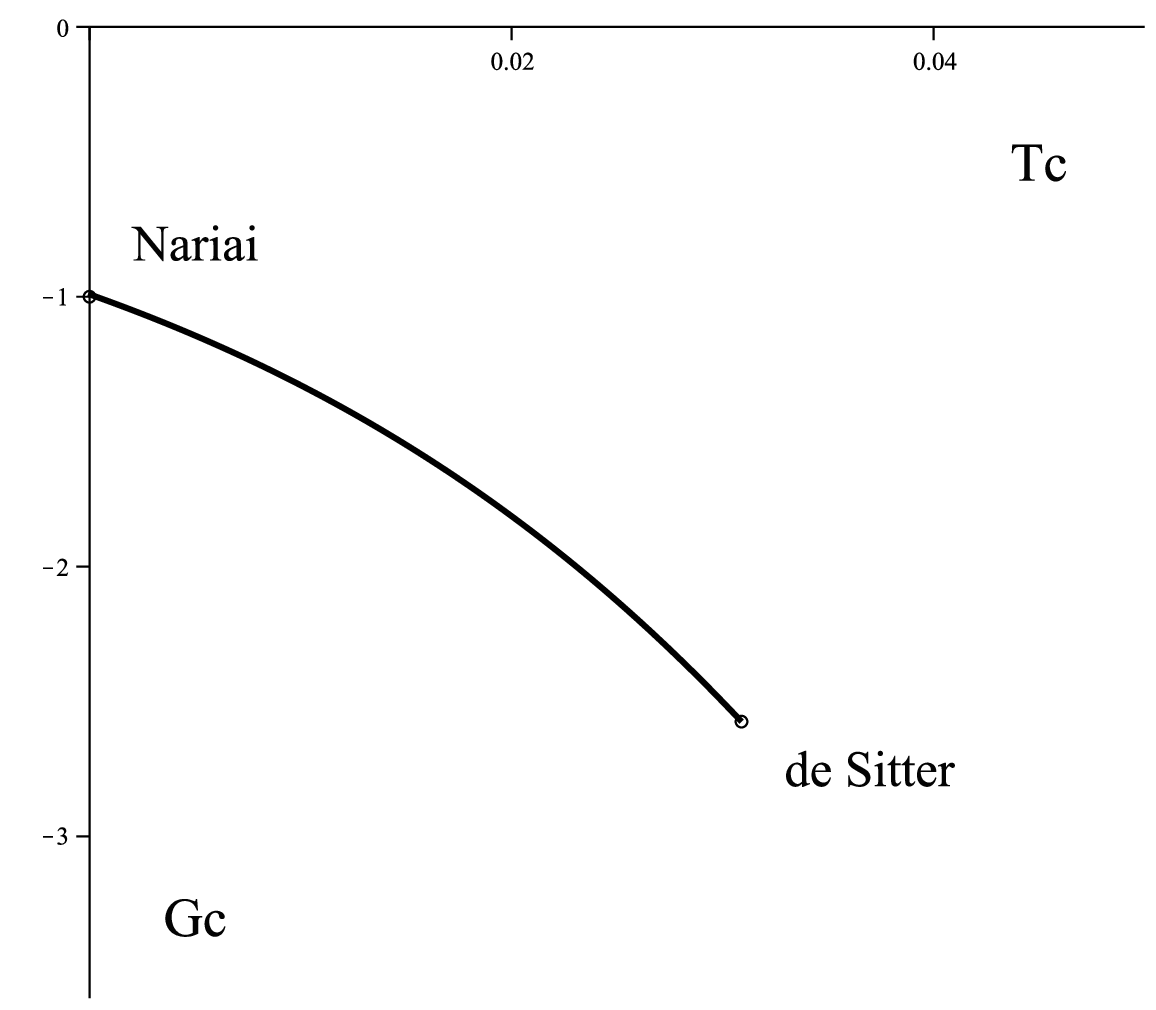}} &
%\rotatebox{-90}{
\includegraphics[width=0.47\textwidth,height=0.27\textheight]{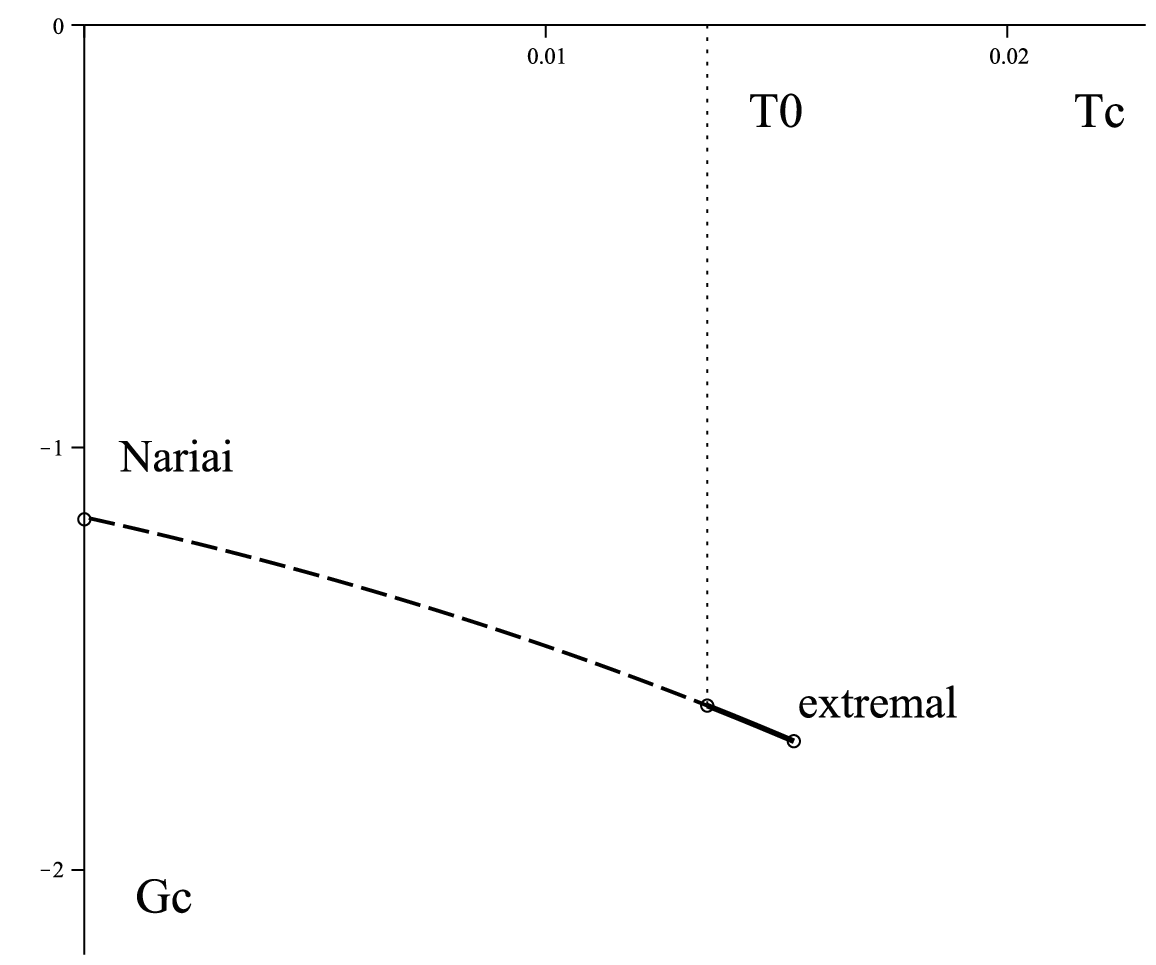}\\
\end{tabular}
\caption{{\bf Gibbs free energy: cosmological horizon.}
The Schwarzschild-dS (left) and the charged dS with $Q=1$ (right) cases are displayed for $P=-0.0045$.
In both instances $G_c$ monotonically decreases and does not indicate the presence of any thermodynamically unstable regions or phase transitions. However,
for the charged case (right) a thermodynamically unstable region, indicated by the dashed black curve, exists for $T_c\in (0, T_0)$. This region corresponds to the upper thermodynamically unstable branch of $G$ for the black hole horizon displayed in Fig.~1; $T_0$ is the temperature of the cosmological horizon that corresponds to the cusp of the black hole horizon Gibbs free energy.
}
\label{fig1c}
\end{figure}

Finally, for the charged dS case the thermodynamic behavior of the {\em inner horizon} should also be taken into account.
This is formally similar to that of the cosmological horizon, with $r_c$ replaced by $r_i$, resulting in the definition
$G_i=G_i(T_i, {P}, Q)=-M-T_{i}S_i$, and is displayed in Fig.~\ref{figInner}.

One can also study the specific heats of the cosmological and inner horizons. These are now defined as
\be
C_c=\frac{\partial H_c}{\partial T_c}\Bigr|_{P,Q}=-\frac{\partial M}{\partial T_c}\Bigr|_{P,Q}\,,\quad
C_i=\frac{\partial H_i}{\partial T_i}\Bigr|_{P,Q}=-\frac{\partial M}{\partial T_i}\Bigr|_{P,Q}\,,
\ee
and displayed for our test spacetimes in Figs.~\ref{fig:CS} and \ref{fig:CRN}. We find that $C_c$ is manifestly positive (apart from the Nariai limit where it vanishes) for both charged and uncharged dS black holes. $C_i$ is only defined for the charged dS case; it is negative (apart from the extremal limit where it vanishes) indicating the local thermodynamic instability of the inner horizon.

\begin{figure}
\centering
\begin{tabular}{cc}
{\includegraphics[width=0.6\textwidth,height=0.3\textheight]{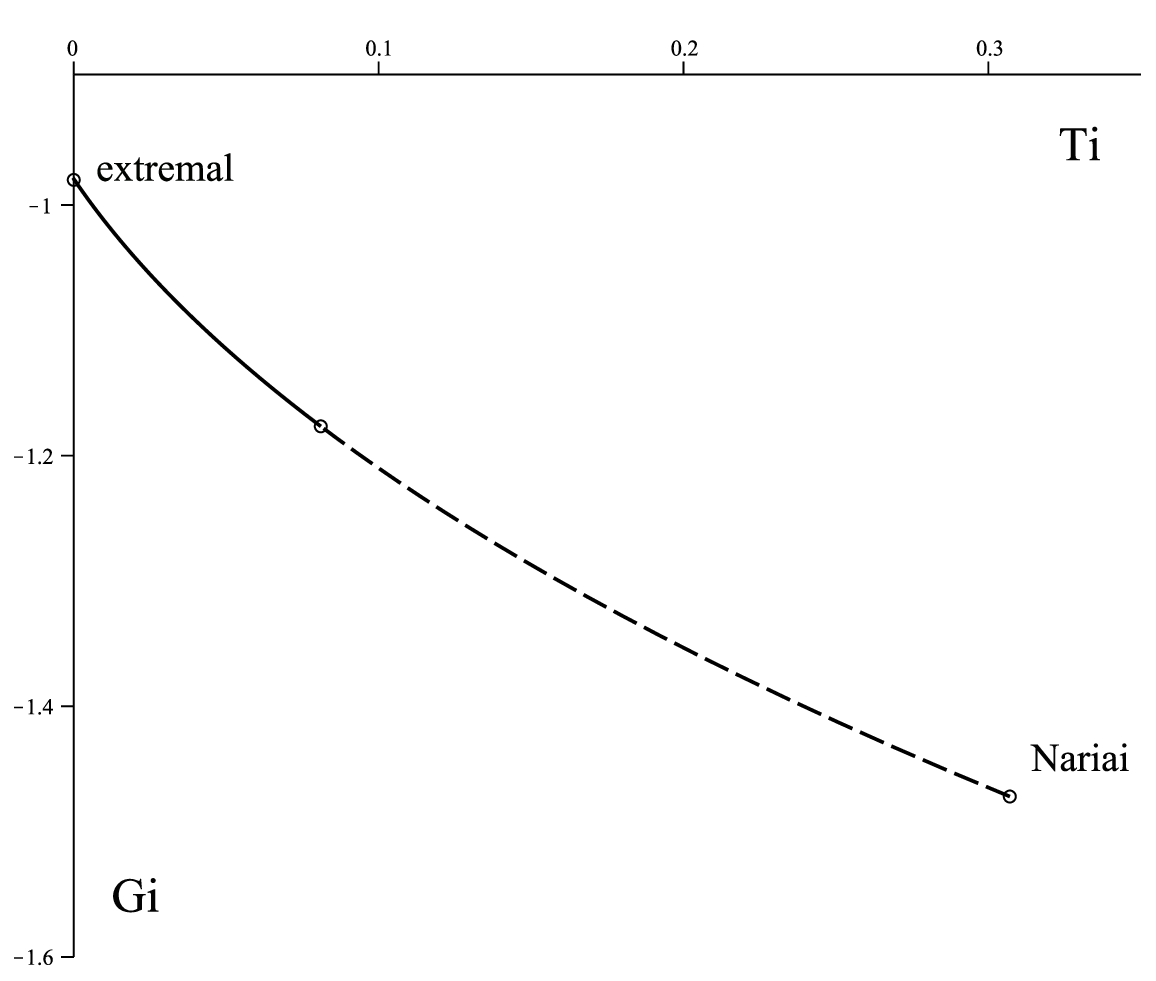}} &
%\rotatebox{-90}{
%\includegraphics[width=0.47\textwidth,height=0.27\textheight]{GcRN.eps}\\
\end{tabular}
\caption{{\bf Gibbs free energy: inner horizon.}
The $G_i-T_i$ diagram is displayed for the charged dS black hole with $Q=1$ and $P=-0.0045$.
}
\label{figInner}
\end{figure}

\subsection{Unified description of all horizons}
To treat all the horizons simultaneously, we note that
\ba
G_c(r_c, {P}, {Q})&=& -G(r_b\to r_c, {P}, {Q})\,,\nonumber\\
T_c(r_c, {P}, {Q})&=&-T_b(r_b\to r_c, {P}, {Q})\,,
\ea
and similarly for the inner horizon. This means that, similar to the AdS case discussed in App.~A, the information about the thermodynamics of all horizons is in fact encoded in $G$, defined by Eq.~\eqref{BHGibbs}, provided we extend its validity to ``{\em all admissible radii}'' $r_b$ and correspondingly to negative
temperatures $T_b$. Such negative temperatures then correspond to the positive temperatures of the cosmological and inner horizons and in these regions the thermodynamic equilibrium occurs for the global maximum (rather than minimum) of $G$.

Specifically, let us consider the charged dS black hole with $Q\neq 0$. The thermodynamics of all three horizons is encoded in $G=G(T,P,Q)$, given by
\be
G=\frac{r}{4}-\frac{2}{3}\pi r^3 P+\frac{3}{4}\frac{Q^2}{r}\,,\quad
T=\frac{1}{4\pi r}+2rP-\frac{Q^2}{4\pi r^3}\,.
\ee
Let us fix $P$ and $Q$ and denote by $r_E$ the corresponding radius of the extremal black hole and by $r_N$ the radius of the Nariai black hole.
We further denote by $r_m$ the minimal radius of the inner horizon, which occurs for the Nariai limit, and by $r_M$ the maximal radius of the cosmological horizon, which occurs for the extremal black hole. We then parametrically plot $G=G(T)$ for the range $r\in(r_m, r_M)$, being careful to interpret the thermodynamic quantities for different ranges of $r$. % in accordance with (2.9).
For $r\in (r_m,r_E)$, $G=-G_i$ and $T=-T_i$ represent inner horizon quantities and the global maximum of $G$ corresponds to a thermodynamically preferred state. For $r\in(r_E,r_N)$, $G=G_b$ and $T=T_b$ represent black hole horizon quantities while $G$ is minimized by the thermodynamically preferred state. For $r\in(r_N, r_M)$, $G=-G_c$ and $T=-T_c$ represent cosmological horizon quantities and the preferred state again corresponds to the maximum of $G$. This is illustrated in right Fig.~\ref{fig1d}.

The case with $Q=0$ is similar, with the exception that there is no longer an inner horizon. Consequently, the range for $r$ is $r\in (0, r_M)$ and we have $r\in(0,r_N)$ for the black hole horizon quantities and $r\in(r_N,r_M)$ for the cosmological horizon quantities, as displayed on the left of Fig.~\ref{fig1d}.

\begin{figure}
\centering
\begin{tabular}{cc}
{\includegraphics[width=0.47\textwidth,height=0.27\textheight]{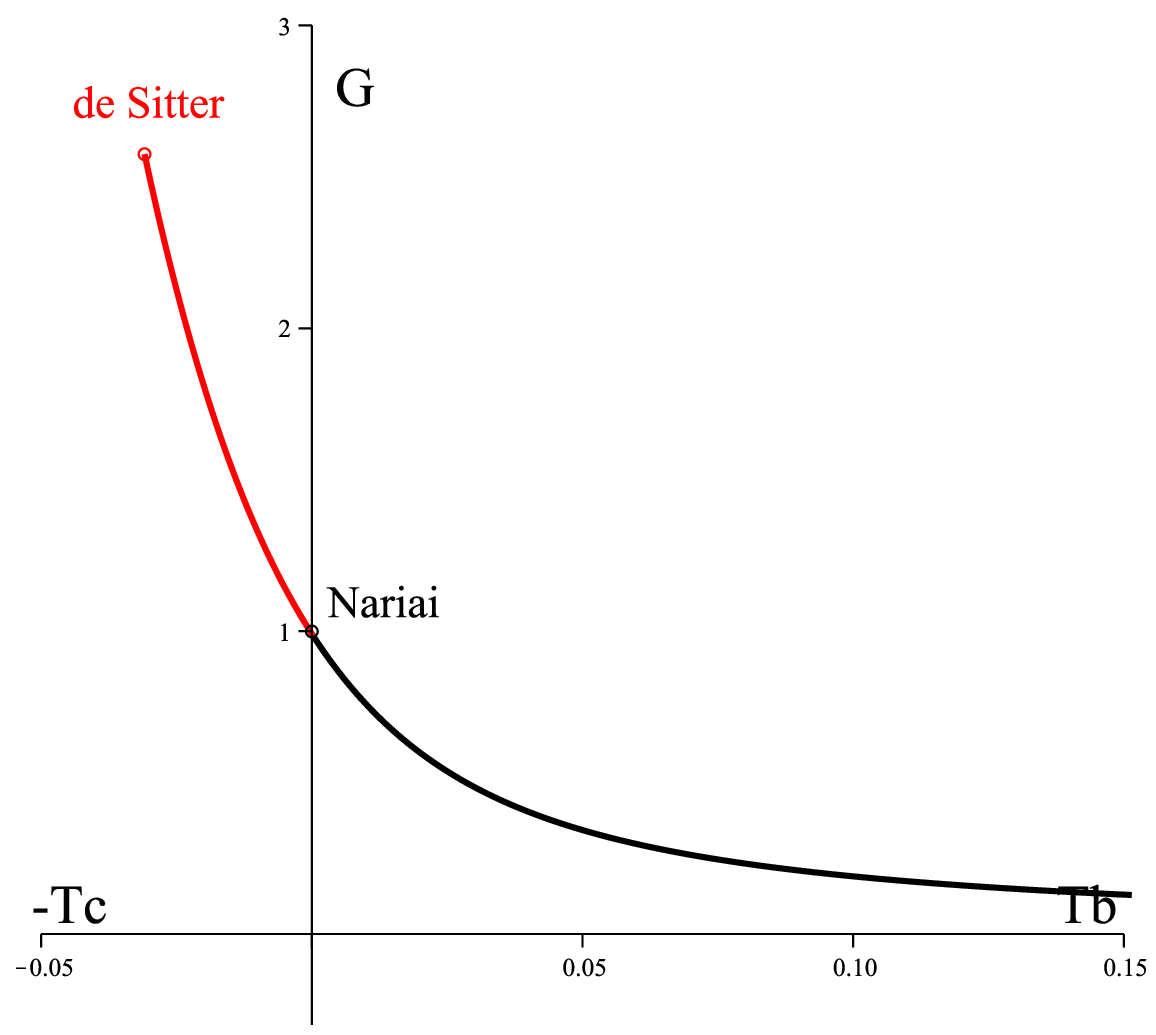}} &
%\rotatebox{-90}{
\includegraphics[width=0.47\textwidth,height=0.27\textheight]{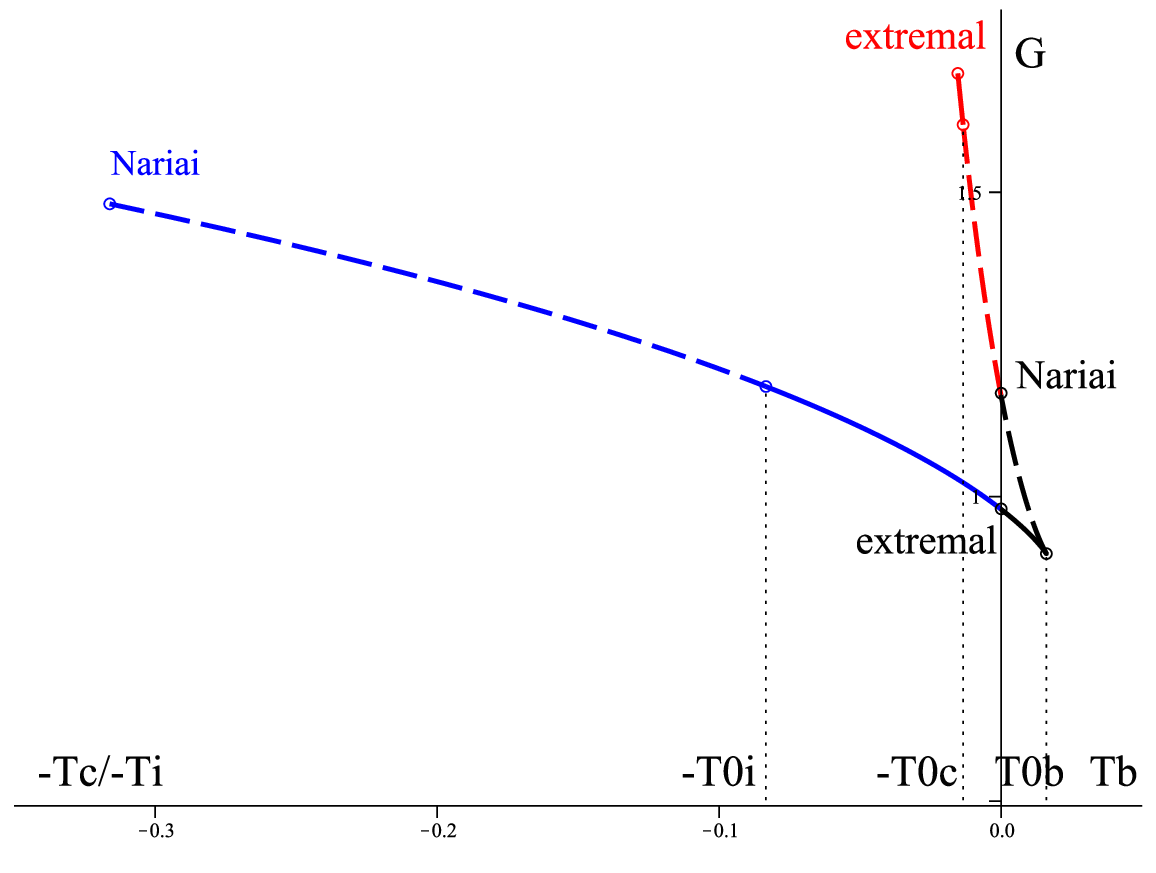}\\
\end{tabular}
\caption{{\bf Gibbs free energy: all horizons.}
The Gibbs free energy $G$ is displayed for all possible horizons: inner (blue), cosmological (red), and black hole (black) for the
Schwarzschild-dS (left) and the charged dS with $Q=1$ (right) cases for $P=-0.0045$; dashed curves correspond to regions that are thermodynamically
unstable.
{Note that the positive $x$-axis corresponds to $T=T_b$ and the negative $x$-axis to $T=-T_c$ or $T=-T_i$ temperatures. For a given state of the black hole this allows one to read temperatures of all horizons. For example, the cusp occurs for the black hole horizon temperature $T_b=T_{b_0}\approx 0.016$; $T_i=T_{i_0}\approx 0.083$ and  $T_c=T_{c_0}\approx 0.013$ are the corresponding temperatures of the inner and cosmological horizons.
}
}
\label{fig1d}
\end{figure}

Having defined the Gibbs free energy $G$ that describes the thermodynamic behavior of {\em all}
involved horizons, we can identify all possible thermodynamically unstable regions and
phase transitions present in the system. For example, consider the charged dS black hole with $Q=1$ and $P=-0.009$ whose thermodynamics is displayed in Fig.~\ref{fig1f}. On the left side we depict the
behavior of the three horizon temperatures, displayed as function of $x=r_b/r_c$. Note that only values of $x$ between extremal (E) and Nariai (N) bounds are admissible, $x\in(x_m=\frac{r_E}{r_N}, 1)$. On the right side the Gibbs free energy is displayed. Let us first concentrate on the black hole horizon.
{Choosing some admissible black hole horizon radius} $r_b=r_{b_0}$ (corresponding to $x=x_0$) in the lower black hole branch, {we can ask what would happen, for example, with the dS system provided we increased} the black hole size $r_b$ a little bit. From the graphs one can immediately read all the temperatures as well as how they are going to change as we increase $r_b$, following the small black arrows. Note also that as long as changes in $r_b$ are small (until we meet the cusp) we move along a lower black hole branch that has lower Gibbs free energy than the upper branch and hence is thermodynamically preferred. However, crossing the cusp would move the system in the upper black hole branch which is thermodynamically unstable. Consequently, we observe unstable regions (dashed) for not only the black hole horizon but also the cosmological (red) and inner (blue) horizon branches. This is how the information about the thermodynamics of each horizon interplays and builds towards the understanding of the thermodynamic behavior of the whole system.
\begin{figure}
\centering
\begin{tabular}{cc}
{\includegraphics[width=0.47\textwidth,height=0.27\textheight]{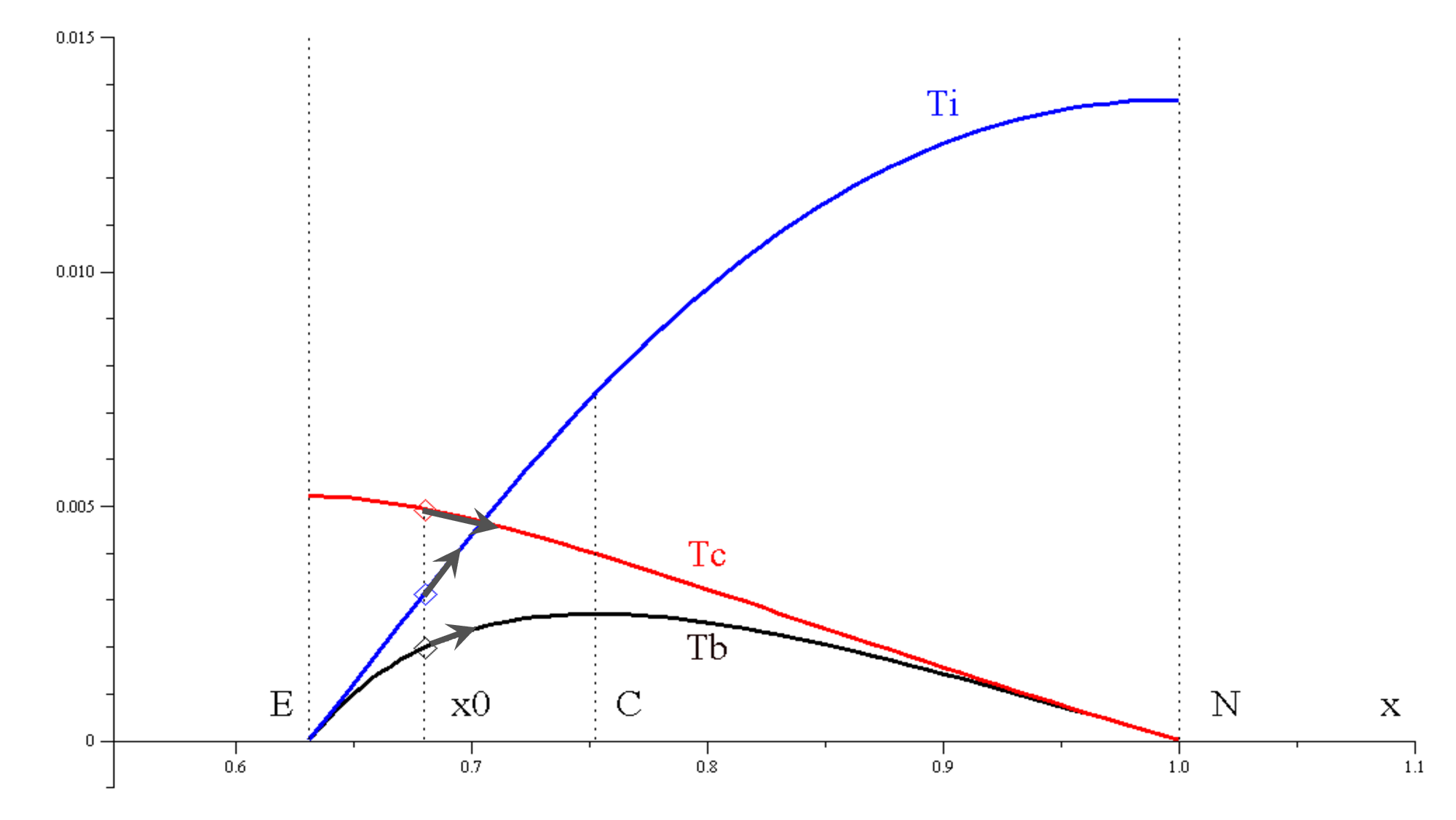}} &
%\rotatebox{-90}{
\includegraphics[width=0.47\textwidth,height=0.27\textheight]{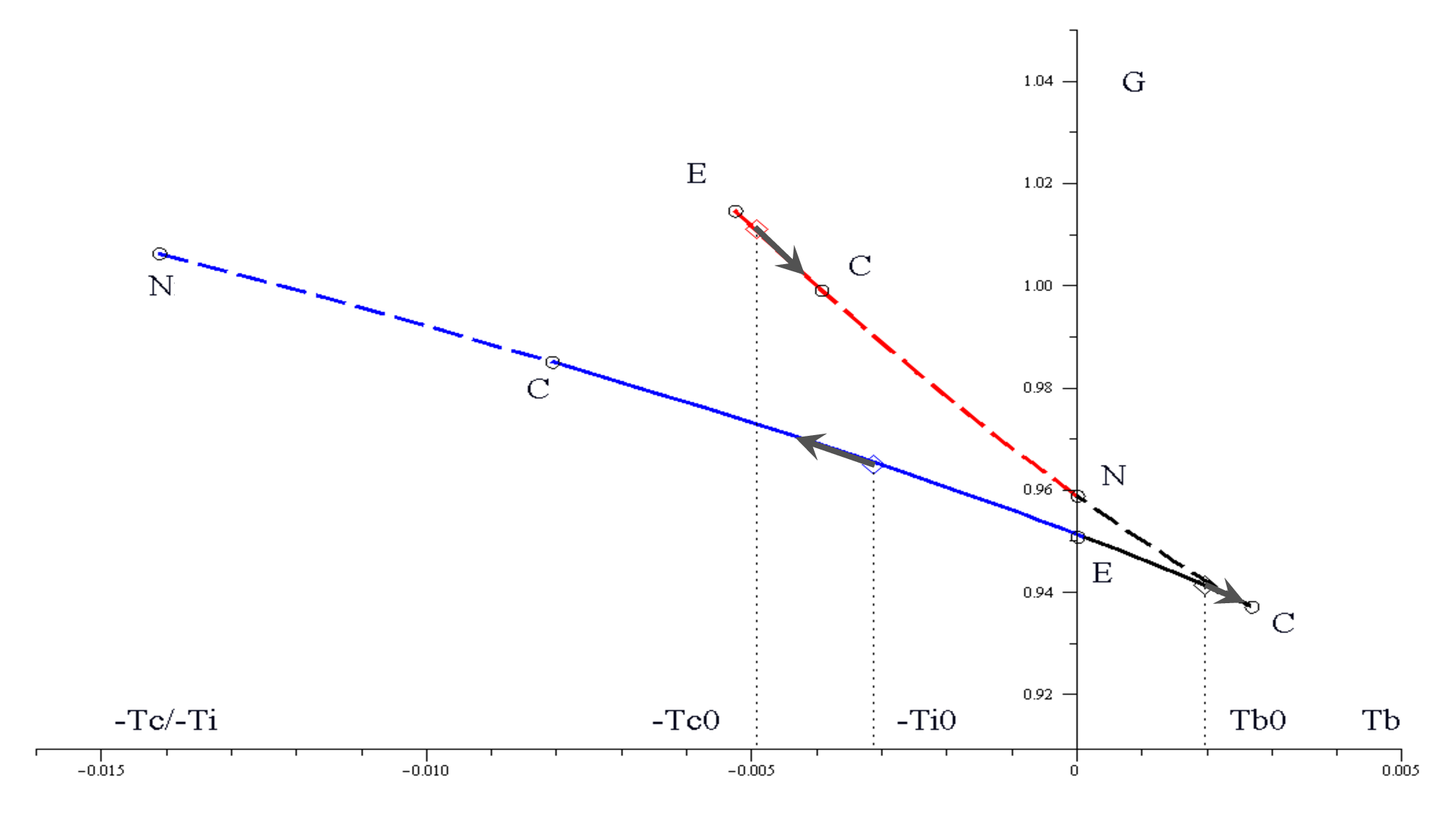}\\
\end{tabular}
\caption{{\bf Thermodynamics of charged dS black holes.}
{\em Left:} The horizon temperatures are displayed as functions of $x=r_b/r_c$. We observe 3 important values of $x$: the extremal limit (E),
the Nariai limit (N), and a cusp (C) for which the black hole horizon temperature reaches its maximum. $x_0$ is chosen as an illustration point; it determines $T_{b_0}, T_{c_0}$ and $T_{i_0}$ for the r.h.s. graph. The arrows illustrate what happens with horizon temperatures as the size of the black hole $r_b$ increases from $r_{b_0}$. Namely, the black hole horizon and inner horizon temperatures locally increase whereas the cosmological horizon temperature decreases. {\em Right:} The corresponding Gibbs free energy, displaying the thermodynamic behavior of all three horizons. In both graphs $P=-0.009$ and $Q=1$.}
\label{fig1f}
\end{figure}

\section{Rotating dS black holes}
In the previous subsection we have studied the thermodynamics of charged dS and Schwarzschild-dS black holes. We observed that both examples are relatively `boring'. Whereas the Gibbs free energy for the Schwarzschild-dS system is completely smooth, the Gibbs free energy of charged dS black hole at least admits a cusp and consequently thermodynamically unstable regions, however no interesting phase transitions are present in either case.
Interesting phase transitions, indicated for example by the presence of a swallow tail, might exist for more complicated dS black hole spacetimes, possibly in higher dimensions. In this section we uncover such behavior by considering  rotating dS black holes in four and higher dimensions.

The general rotating dS metrics in all dimensions and the corresponding thermodynamic charges are gathered in App.~\ref{kerrds}.
Similar to the charged case, we have up to three horizons and the corresponding Nariai and extremal black hole limits.
To investigate the thermodynamic behavior we study the Gibbs free energy $G=G(T,P,J^i)$ in the canonical (fixed $J^i$) ensemble. The corresponding expression
is slightly different for odd and even dimensions, see App.~\ref{kerrds}. For example, in even dimensions we have
\ba
G&=&\frac{\omega_{d-2}}{8\pi}\prod_{k=1}^N \frac{r^2+a_k^2}{\Xi_k}\Bigl[\frac{1-r^2/l^2}{r}\Bigl(\sum_{i=1}^N\frac{1}{\Xi_i}-\frac{r^2}{r^2+a_i^2}\Bigr)-\frac{1+r^2/l^2}{2r}\Bigr]\,,\nonumber\\
T&=&\frac{r(1-r^2/l^2)}{2\pi}\sum_{i=1}^N\frac{1}{r^2+a_i^2}-\frac{1+r^2/l^2}{4\pi r}\,,
\ea
where $N=2N+2$, and $a_i$ are understood as functions of the angular momenta, $a_i=a_i(J_j)$ through formulae \eqref{Ji} and \eqref{ff}.

\subsection{Behavior in four and five dimensions}
It turns out that (similar to the AdS case) the behavior of the Gibbs free energy of 4d and 5d rotating dS black holes is qualitatively similar to the behavior of $G$ of the four-dimensional charged-dS black hole. Namely, we observe no interesting features apart from a cusp of divergent specific heat.  This is displayed in Fig.~\ref{fig4dKerr}. In the five-dimensional case we have checked that the same qualitative behavior is obtained for various ratios $q=J_1/J_2$ of the two angular momenta. As we shall see in the next subsection, a new feature appears in six dimensions.

\begin{figure}[!htb]
\centering
\begin{tabular}{cc}
{\includegraphics[width=0.47\textwidth,height=0.27\textheight]{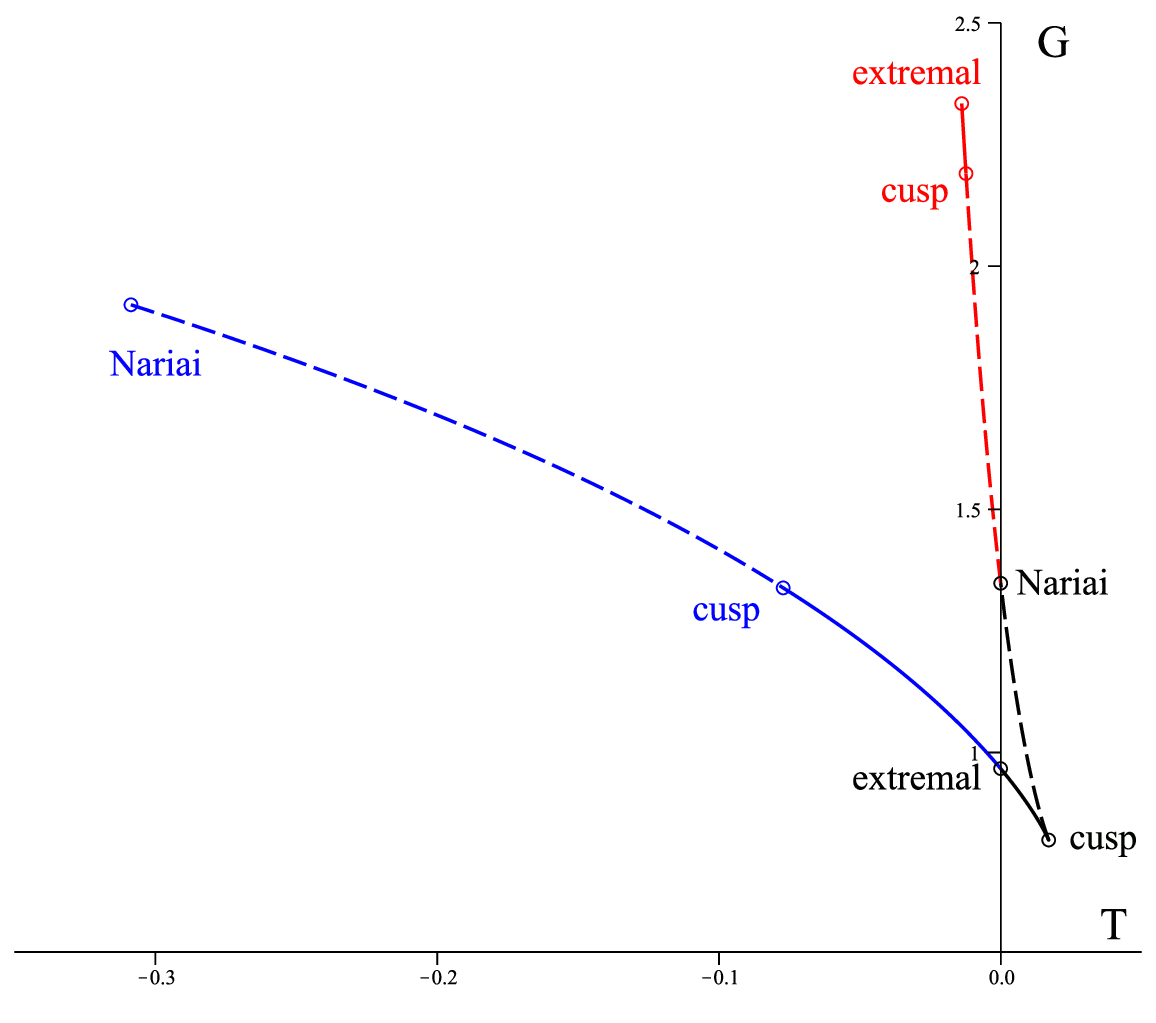}} &
%\rotatebox{-90}{
\includegraphics[width=0.47\textwidth,height=0.27\textheight]{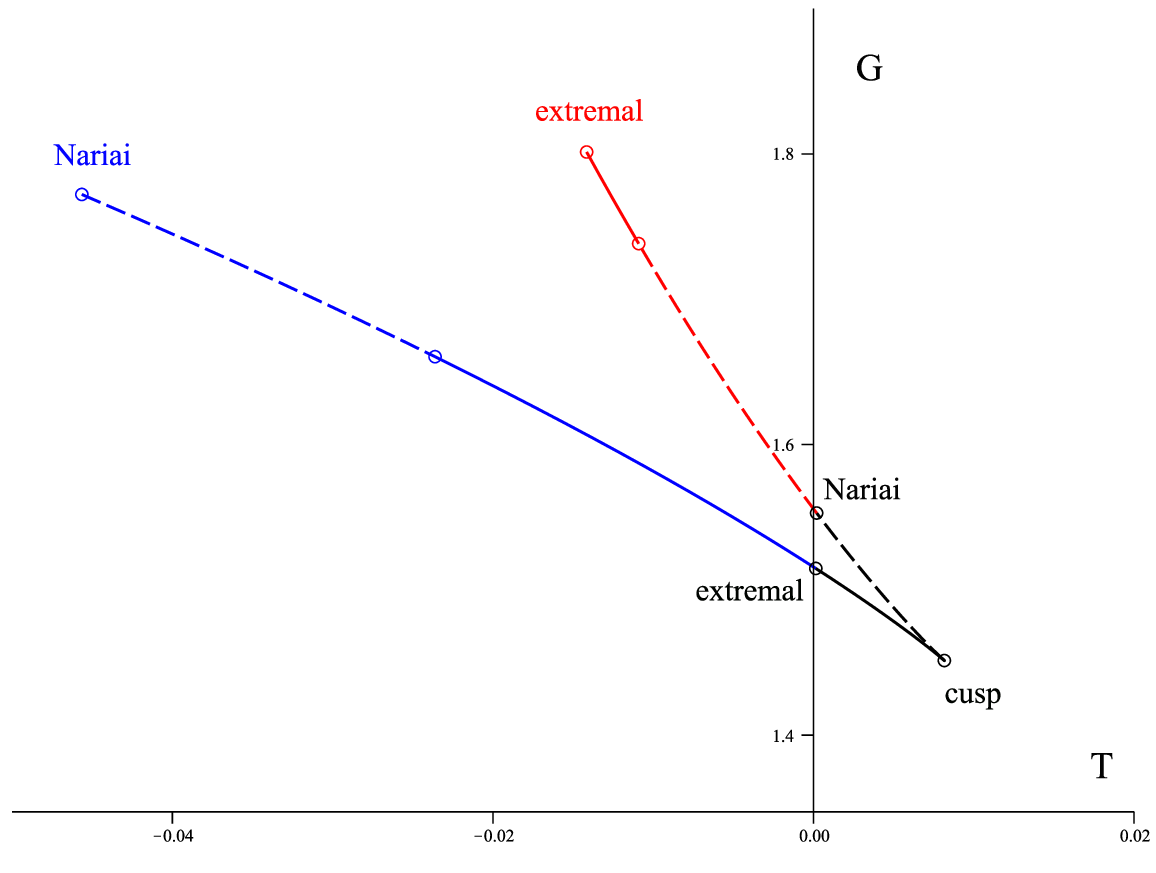}\\
\end{tabular}
\caption{{\bf Thermodynamics of rotating dS black holes in four and five dimensions.}
{\em Left} figure displays the $G-T$ diagram for the four-dimensional Kerr-dS black hole for $P=-0.0025$ and $J=1$.
{\em Right} figure displays the $G-T$ diagram for a doubly-spinning Kerr-dS black hole in five dimensions for $P=-0.05$ and $q=J_1/J_2=0.5$.
In both instances we observe qualitatively the same behavior as in the 4-dimensional charged-dS case.
}
\label{fig4dKerr}
\end{figure}

\subsection{Six dimensions: reentrant phase transitions}
In the six-dimensional case the behavior of $G$ becomes more interesting.
Namely, in Fig.~\ref{fig1b} we display the Gibbs free energy for the
doubly spinning Kerr-dS black hole in $d=6$ dimensions for  fixed angular momentum ratio $q=J_1/J_2=0.112$.\footnote{%
As with the AdS case, the particular behavior of $G$ depends crucially on the choice of ratio $J_1/J_2$. For example, for a given $q=0.112$, we can find the reentrant phase transition in the range of pressures $P\in(0,P_c)$ where $P_c\approx -0.0022$.}
The complicated behavior of $G$ (in the black hole horizon region) indicates the presence of a {\em reentrant phase transition}, exactly analogous to the one observed for AdS black holes in \cite{Altamirano:2013ane}. Specifically, as the black hole horizon temperature $T_b$ monotonically increases, we observe a reentrance of the small black hole phase as follows. Below $T_b=T_1$ a (lower left) branch of small black holes is preferred, at $T_1$ a first order phase transition occurs and
the (lower middle) large black hole phase becomes temporarily dominant. This continues until at
$T_b=T_0$ the large black hole branch terminates. Increasing the temperature even further results in a zeroth order phase transition as the systems `jumps back' to the (lower right) branch of small black holes. {One can easily check that in this case the branches with lower Gibbs (for a given temperature) are locally thermodynamically stable as they have positive specific heat.}
In other words, we observe a small/large/small black hole reentrant phase transition, similar to the AdS case.\footnote{Note, however, that in the AdS case the reentrant phase transition occurs for singly spinning Kerr-AdS black holes in six dimensions, whereas the doubly spinning ones may feature a tricritical point and small/intermediate/large black hole phase transition \cite{Altamirano:2013uqa}.
}

\begin{figure}
\centering
%\rotatebox{-90}{
\includegraphics[width=0.75\textwidth,height=0.33\textheight]{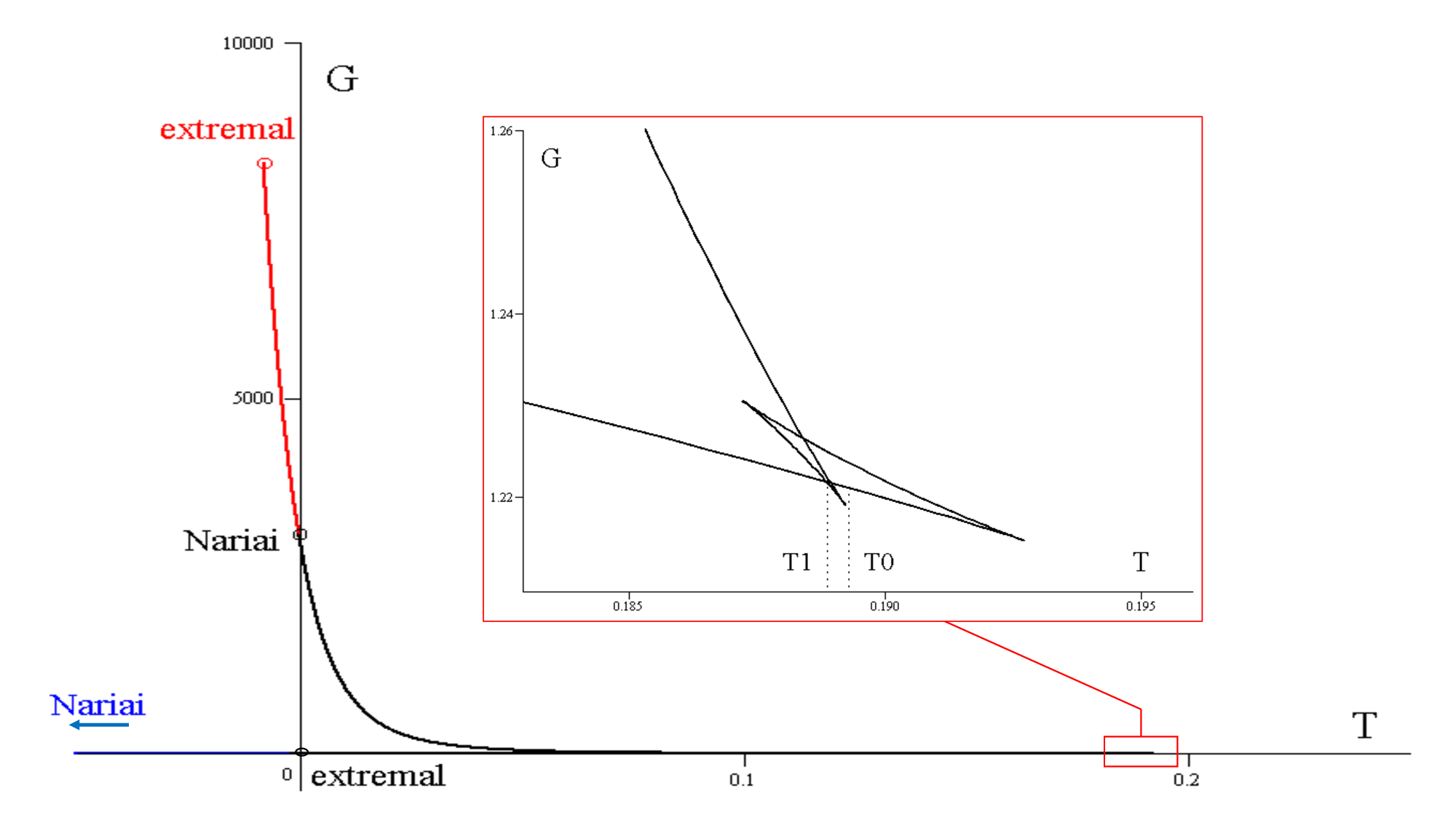}
\caption{{\bf Reentrant phase transition.}
The black hole horizon Gibbs free energy is displayed for the doubly spinning Kerr-dS black hole in $d=6$ dimensions for fixed angular momentum ratio $q=J_1/J_2=0.112$ and $P=-0.001$. %The black curve corresponds to the dS case with $P=-0.001$ whereas the red curve to the AdS case with $P=0.003$. In
%both instances we observe a reentrant phase transition.
We observe a typical behavior characteristic of the reentrant phase transition, previously observed for the AdS black holes \cite{Altamirano:2013uqa}.
Namely, as the temperature increases the system undergoes small/large/small black hole reenrant phase transition: as the temperature increases, the first order phase transition from small to large black hole at $T=T_1$ is followed by a zeroth order phase transition at $T=T_0$ as the system `jumps back' to the small black hole branch.}
\label{fig1b}
\end{figure}

\section{Conclusions}

The subject of black hole thermodynamics continues to be one of great importance in gravitational physics. Over the past fifty years many of the studies in this field have concentrated on the thermodynamics of asymptotically flat and asymptotically anti de Sitter black holes, the latter emerging in light of the AdS/CFT correspondence. However, the thermodynamics of asymptotically de Sitter black holes possesses new difficulties and remains relatively unexplored. One of the primary reasons is that de Sitter black hole spacetimes are essentially non-equilibrium multi-temperature
systems; each horizon %can be assigned
is characterized by
its own temperature and a first law-like relation. Although such laws are not independent and various quantities in them are correlated, it is by no means obvious, despite several attempts in the literature, that a reasonable equilibrium thermodynamic description based on the effective temperature approach can be consistently formulated. This however does not mean that one has to give up on the thermodynamics of these systems.

In this paper we proposed an alternative strategy for studying the thermodynamics of de Sitter black holes. %, that is in a way remarkably analogous to the AdS systems.
Namely, we formulated several thermodynamic first laws (one for each horizon, as usual) and studied their thermodynamics as if
they were independent thermodynamic systems characterized by their own temperature.
The internal correlation of these systems then enabled us to
define a single Gibbs free energy-like quantity $G$  that captures their individual thermodynamic behavior, as well as containing information about possible phase transitions of the whole de Sitter system once properly combined.
In this way we were able to uncover the rich phase structure of de Sitter black holes. Namely, although we found that the thermodynamic behavior of the four-dimensional Schwarzschild-dS and charged dS black holes is relatively uninteresting, we discovered the presence of reentrant phase transitions for Kerr-dS black holes in six dimensions.

{As a simple consistency check we also defined and investigated the horizon specific heats.
In particular, we found that for the studied cases whenever a possibility of multiple black hole branches emerges (for a given temperature) the branch with preferred Gibbs has also positive specific heat and is locally thermodynamically stable.}
Is our alternative proposal for studying the thermodynamics of dS black holes internally self-consistent? And if so, how many of the phase transitions and critical phenomena observed for the AdS black holes find their analogue in the dS case? These remain  interesting open questions left for future investigations.

\section*{Acknowledgements}
%We would like to thank D.~Kastor and J.~Traschen for discussions in  the early stage of this work.
We would like to thank R.~Mann for reading the manuscript and useful comments and D.~Kastor and J.~Traschen for discussions in  the early stage of this work. {We are also grateful to the anonymous referee for his/her comments leading to a significant improvement of the manuscript.}
This research was supported in part by the Perimeter Institute for Theoretical Physics and by the Natural Sciences and Engineering Research Council of Canada. Research at the Perimeter Institute is supported by the Government of Canada through Industry Canada and by the Province of Ontario through the Ministry of Research and Innovation.

\appendix

\section{AdS analogue of liquid/gas phase transition: inner horizon}\label{appA}
{
In this appendix we re-investigate the thermodynamic analogue of a liquid/gas phase transition present in the charged AdS black hole spacetime  \cite{Chamblin:1999tk, Chamblin:1999hg, Cvetic:1999ne, Kubiznak:2012wp}. The purpose is twofold.
i) To show that a ``multi-horizon description'' similar to the one studied in the main
text is already (silently) present in the AdS case.
ii) To address the following question overlooked in the existing literature: What happens
to the inner horizon when the outer horizon experiences a Van der Waals-like phase transition?
}

\begin{figure}
\centering
\begin{tabular}{cc}
{\includegraphics[width=0.47\textwidth,height=0.27\textheight]{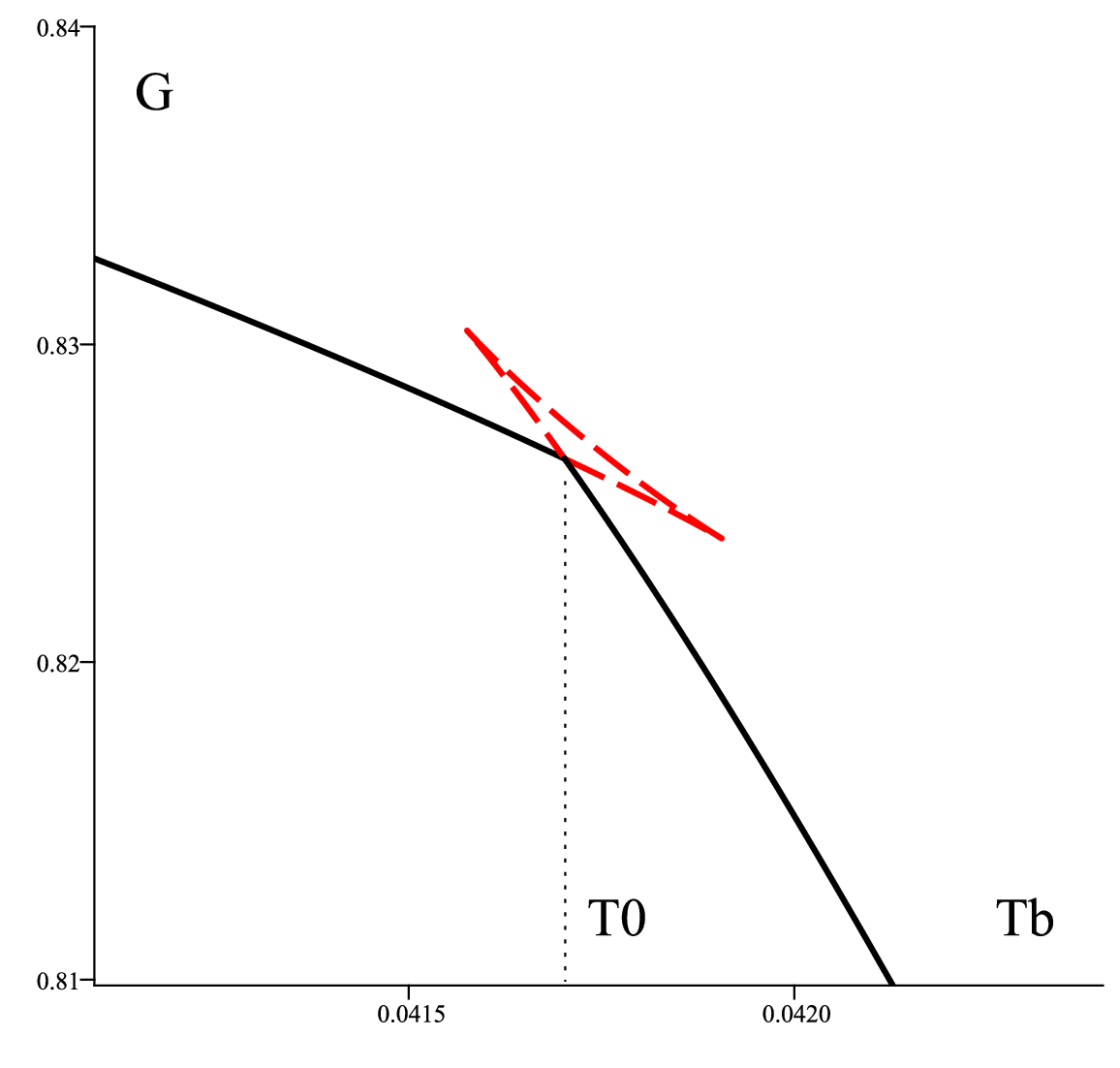}} &
%\rotatebox{-90}{
\includegraphics[width=0.47\textwidth,height=0.27\textheight]{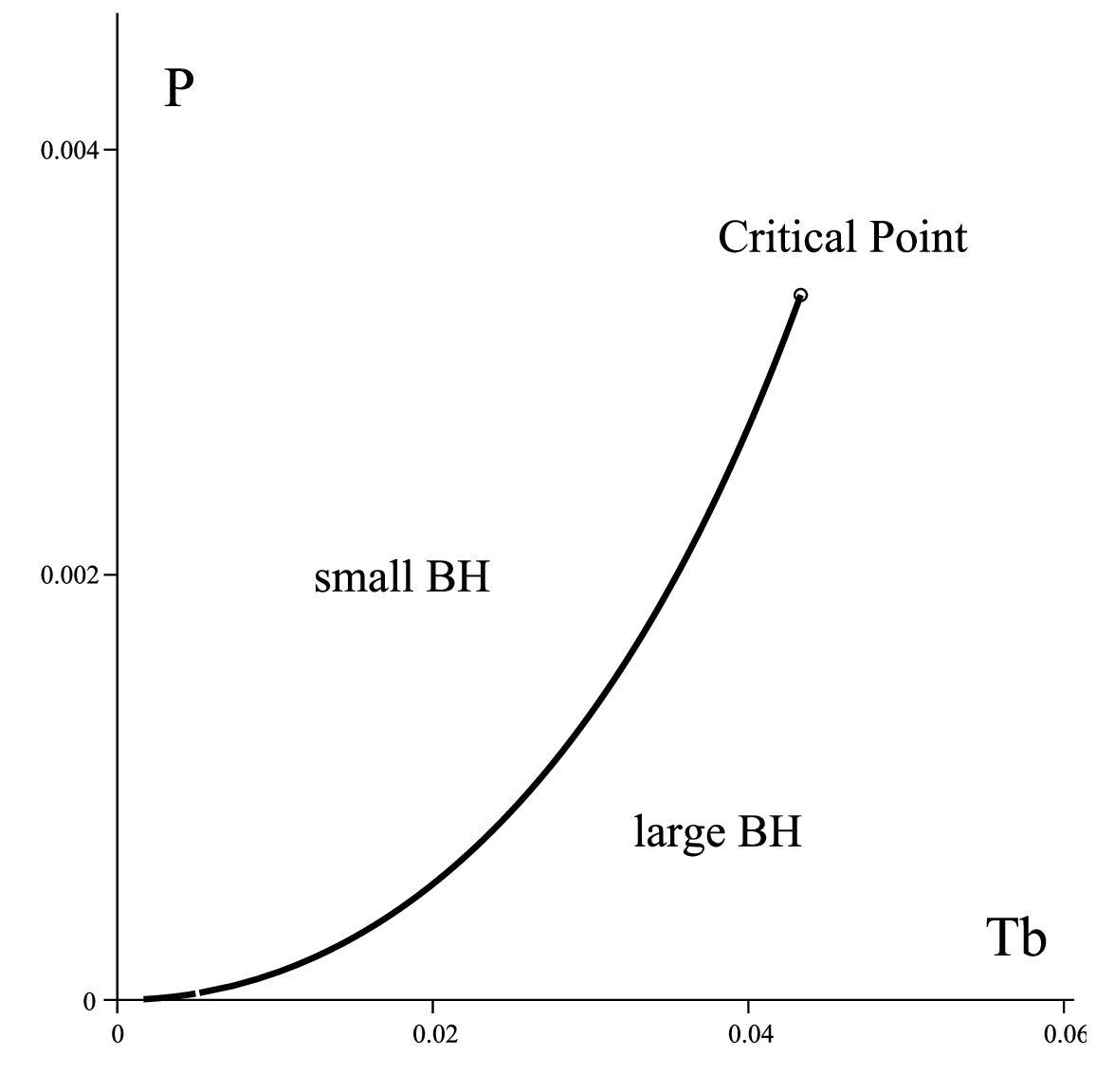}\\
\end{tabular}
\caption{{\bf AdS analogue of liquid/gas phase transition.}
{\em Left:} The Gibbs free energy $G$  is displayed for the charged AdS black hole for $Q=1$ and $P=0.003$; we observe the characteristic swallow tail behavior: the thermodynamically unstable configuration of the outer black hole horizon is displayed by the dashed red curve, temperature $T_b=T_0$ corresponds to the first-order phase transition. {\em Right:} The corresponding $P-T_b$ phase diagram is reminiscent of the liquid/gas phase transition.
}
\label{Appfig1}
\end{figure}
To this end, we consider a charged asymptotically AdS black hole in four dimensions, given by the metric
\be\label{RNmetric}
ds^2=-fdt^2+\dfrac{dr^2}{f}+r^2d\Omega_2^2\,,\quad f=1-\dfrac{2M}{r}+\frac{Q^2}{r^2}+\frac{r^2}{l^2}\,,
\ee
and identify the negative cosmological constant with positive pressure, according to $P=(d-1)(d-2)/(16\pi l^2)$.
The black hole horizon is located at the largest $r_b>0$ for which $f(r_b)=0$. The thermodynamic quantities then read
\be
T_b=\dfrac{1}{4\pi r_b}\Big(1+8\pi Pr_b^2-\dfrac{Q^2}{r_b^2}\Big)\,,\quad
S_b=\pi r_b^2\,,\quad \Phi_b=\dfrac{Q}{r_b}\,,\quad V_b=\dfrac{4}{3}\pi r_b^3\,,\quad
\ee
and obey the first law of black hole thermodynamics
\be
dM=T_bdS_b+\Phi_b dQ+V_b dP\,.\label{1aAdS}
\ee
For sufficiently small pressure ${0<P<P_c=1/96\pi Q^2}$, the  Gibbs free energy
\be\label{appG}
G=M-T_bS_b.
\ee
possesses a swallow tail, which represents the gravitational analogue of a liguid/gas phase diagram, see Fig.~\ref{Appfig1}. Namely, there is a first order small black hole/large black hole phase transition. During this transition, the radius of the black hole $r_b$ suffers from a jump while the temperature $T_b$ and pressure $P$ remain constant.

\begin{figure}
\centering
\begin{tabular}{c}
{\includegraphics[width=0.5\textwidth,height=0.25\textheight]{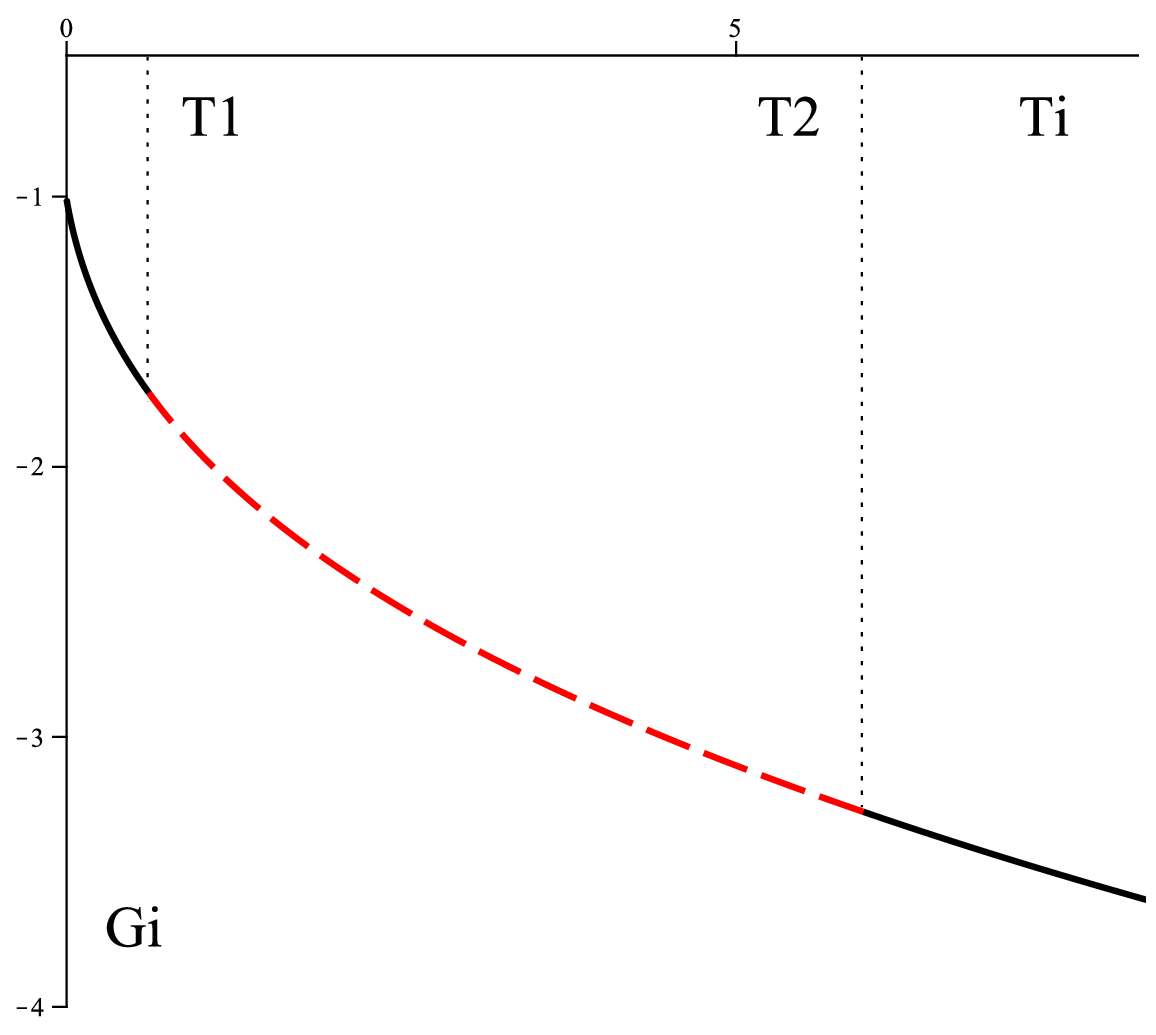}} \\
\end{tabular}
\caption{{\bf Behavior of $G_i$.}
The behavior of the inner horizon Gibbs free energy $G_i$ is displayed for $P=0.003$ and $Q=1$. The two black curves correspond to thermodynamically stable inner horizon configurations whereas the dashed red curve in between $T_1$ and $T_2$ corresponds to the thermodynamically unstable configuration of the inner horizon for which the outer horizon maintains temperature $T_0$ and undergoes a first order phase transition.
}
\label{figApp2}
\end{figure}
However, the black hole spacetime also possesses an inner black hole horizon, located at $r_i$, $0\leq r_i\leq r_b$, $f(r_i)=0$.
What happens to this horizon during the above phase transition?
It is straightforward to show that the inner horizon can be assigned the following thermodynamic quantities:
\be
T_i=\dfrac{-1}{4\pi r_i}\Big(1+8\pi P r_i^2-\dfrac{Q^2}{r_i^2}\Big)\,,\quad
S_i=\pi r_i^2\,,\quad \Phi_i=\dfrac{Q}{r_i}\,,\quad V_i=\dfrac{4}{3}\pi r_i^3\,,\quad
\ee
that formally satisfy the following first law:
\be
dM=-T_idS_i+\Phi_i dQ+V_i dP\,,\label{1bAdS}
\ee
Identifying the enthalpy $H_i=-M$ we can write this as
$dH_i=T_idS_i-V_i d{P}-\Phi_i d{Q}$ (note the inevitable minus sign in the work terms) and hence identify the corresponding Gibbs free energy as
\be
G_i=-M-T_iS_i\,.
\ee
The behavior of $G_i$ is displayed in Fig.~\ref{figApp2}. We observe a monotonic behavior seemingly without any phase transitions. This is misleading however. It is easy to show that while the black hole undergoes a first order phase transition and the horizon radius $r_b$ jumps, the inner horizon $r_i$ jumps as well. Consequently, the inner horizon temperature $T_i$ (which is a monotonic function of $r_i$) also jumps from $T_1$ to $T_2$, unlike the black hole horizon temperature $T_b=T_0$ which remains constant. This means that in fact only a part of the displayed Gibbs free energy $G_i$ corresponds to a stable equilibrium state and the region in between $T_1$ and $T_2$ is thermodynamically unstable.
In other words, the black hole first order phase transition leaves an imprint on the inner horizon: the inner horizon suffers from both a jump in radius and a jump in temperature.
%The Similar conclusions also apply to the dS and inner horizons of the  asymptotically dS spacetime studied in the main text, see Sec.~3.

%Although the inner horizon Gibbs free energy seems trivial, the black hole first order phase transition has its imprint on the inner horizon: the inner horizon suffers from both a jump in radius and a jump in temperature. Similar conclusions also apply to the dS and inner horizons of the  asymptotically dS spacetime, during the possible first order black hole phase transitions, see Sec.~3.

\begin{figure}
\centering
\begin{tabular}{c}
{\includegraphics[width=0.6\textwidth,height=0.27\textheight]{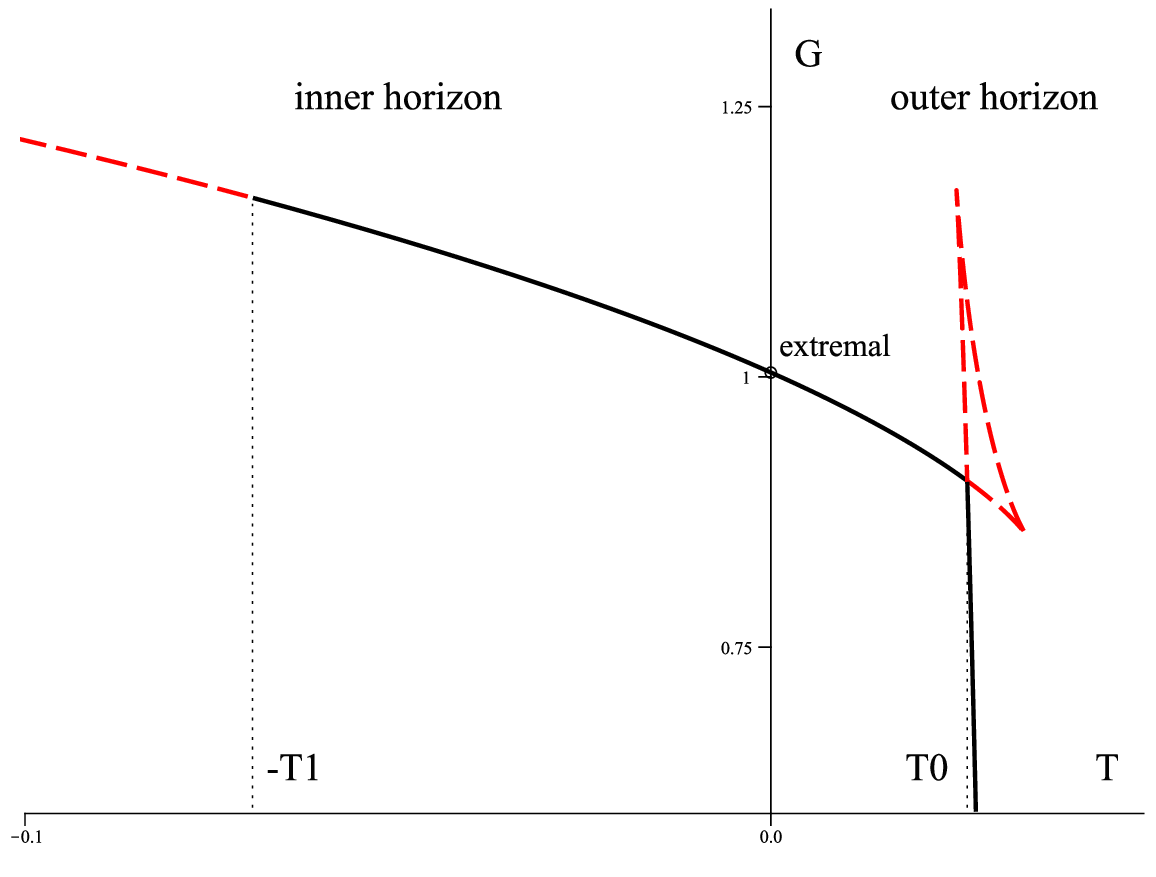}} \\
\end{tabular}
\caption{{\bf Thermodynamic description of both horizons.}
The Gibbs free energy $G$ is displayed for both positive and negative temperatures to capture the thermodynamic behavior of the inner and outer horizons simultaneously. We have set $Q=1$ and $P=0.001$; for such a choice $T_2$ becomes too large to be displayed in this figure.
}
\label{figApp3}
\end{figure}
Let us finally remark that we can write
\be
G_i(r_i,P,Q)=-G(r_b\to r_i,P,Q)\,,\quad T_i(r_i,P,Q)=-T_b(r_b\to r_i,P,Q)\,.
\ee
For this reason we may formally obtain the thermodynamic information about both horizons by studying $G$, \eqref{appG}, extended to
all admissible non-negative radii and hence formally to negative temperatures. Negative temperature corresponds to the positive inner horizon temperature, while in such a region the thermodynamic equilibrium state maximizes (rather than minimizes) $G$. This is displayed in Fig.~\ref{figApp3}.

%%%%%%%%%%%%%%%%%%%%%%%%%%%%%%%%%%%%%%%%%%%%%%%%%%%%%%%%%%%%%%%%%%%%%%%%%%%%%%%%%%%%%%%%
%%%%%%%%%%%%%%%%%%%%%%%%%%%%%%%%%%%%%%%%%%%%%%%%%%%%%%%%%%%%%%%%%%%%%%%%%%%%%%%%%%%%%%%%
\section{Rotating de Sitter black holes and their thermodynamics}\label{kerrds}
In this appendix we review $d$-dimensional Kerr-de Sitter black holes and their thermodynamics \cite{Gibbons:2004uw, Gibbons:2004js, Gibbons:2004ai, Dolan:2013ft}. The metric solves the vacuum Einstein equations with cosmological constant $\Lambda$, \eqref{P}, and reads
\begin{eqnarray}\n{MPC}
ds^2&=&-W(1-r^2/l^2)d t^2+\frac{2m}{U} \Bigl(Wd t-
\sum_{k=1}^{N}\frac{a_k\mu_k^2d \varphi_k}{\Xi_k}\Bigr)^{2}
\nonumber\\
&+&\sum_{k=1}^{N}\frac{r^2+a_k^2}{\Xi_k}\,(\mu_k^2d \varphi_k^2
+d \mu_k^2)+\frac{Ud r^2}{f}+\epsilon r^2 d \nu^2\nonumber\\
&+&
\frac{1/l^2}{W(1-r^2/l^2)}\,
\Bigl(\sum_{k=1}^N\frac{r^2\!+\!a_k^2}{\Xi_k}\,\mu_k d \mu_k\!+\!
\epsilon r^2\nu d \nu\Bigr)^2,\ \
\end{eqnarray}
where
\begin{eqnarray}\label{ff}
W&=&\sum_{k=1}^N\frac{\mu_k^2}{\Xi_k}+\epsilon \nu^2\,,\quad
f=r^{\epsilon-2}(1 -r^2/l^2)\prod_{k=1}^{N}(r^2+a_k^2)-2m\,,\nonumber\\
U&=&r^\epsilon\Bigl(\sum_{i=1}^{N}\frac{\mu_i^2}{r^2+a_i^2}+\frac{\epsilon \nu^2}{r^2}\Bigr)\prod_{k=1}^N(r^2+a_k^2)\,,\ \,\Xi_k=1+\frac{a_k^2}{l^2}\,.\ \
\end{eqnarray}
Here $N\equiv [(d-1)/2]$, where $[A]$ means the integer part of $A$, and we
have defined $\epsilon$ to be $1$ for even $d$ and $0$ for odd $d$. The
coordinates $\mu_i$ are not independent, but obey the constraint
\begin{equation}\label{constraint}
\sum_{k=1}^N\mu_{k}^2+\epsilon \nu^2=1\,.
\end{equation}
We have up to three horizons (two black hole horizons and a cosmological dS horizon). These occur at positive radii determined by $f(r)=0$.

%%%%%%%%%%%%%%%%%%%%%%%%%%%%%%%%%%%%%%%%%%%%%%%%
\subsection{Even dimensions}
In even dimensions ($d=2N+2$), we find the following thermodynamic quantities \cite{Gibbons:2004ai, Dolan:2013ft}.
The `mass' and angular momenta read
\be\label{Ji}
M=\frac{m \cA_{d-2}}{4\pi \prod_j \Xi_j}\sum_i \frac{1}{\Xi_i}\,,\quad
J_i = \fft{m a_i \cA_{d-2}}{4\pi \Xi_i \prod_j\Xi_j}\,,
\ee
where ${\cal \omega}_{d-2}$ stands for the volume of the unit-radius $(d-2)$-sphere,
\be
{\cal \omega}_{d-2} = \fft{2\pi^{(d-1)/2}}{\Gamma[(d-1)/2]}\,.
\ee
The cosmological horizon is assigned the following entropy, temperature, angular velocities, and thermodynamic volume:
\ba\label{TDevenH}
S_c&=& \ft14 \cA_{d-2}\, \prod_k \fft{r_c^2+a_k^2}{\Xi_k}=\frac{A_c}{4}\,,\quad
T_c= -\fft{r_c(1-r_c^2/l^2)}{2\pi} \sum_k \fft1{r_c^2+a_k^2} +
  \fft{1+r_c^2/l^2}{4\pi r_c}\,,\nonumber\\
\Omega_c^k &=& \fft{(1-r_c^2/l^2) a_k}{r_c^2 + a_k^2}\,,\quad
V_c = \frac{r_c A_c}{d-1}\left[1+\frac{1-r_c^2/l^2}{(d-2)r_c^2}\sum_k \frac{a_k^2}{\Xi_k}\right]\,.
\ea
Similarly, for the black hole horizon and the inner horizon we have the following quantities:
\ba\label{TDevenH}
S_b&=& \ft14 \cA_{d-2}\, \prod_k \fft{r_b^2+a_k^2}{\Xi_k}=\frac{A_b}{4}\,,\quad
T_b= \fft{r_b(1-r_b^2/l^2)}{2\pi} \sum_k \fft1{r_b^2+a_k^2} -
  \fft{1+r_b^2/l^2}{4\pi r_b}\,,\nonumber\\
\Omega_b^k &=& \fft{(1-r_b^2/l^2) a_k}{r_b^2 + a_k^2}\,,\quad
V_b = \frac{r_b A_b}{d-1}\left[1+\frac{1-r_b^2/l^2}{(d-2)r_b^2}\sum_k \frac{a_k^2}{\Xi_k}\right]\,,
\ea
and
\ba\label{TDevenH}
S_i&=& \ft14 \cA_{d-2}\, \prod_k \fft{r_i^2+a_k^2}{\Xi_k}=\frac{A_i}{4}\,,\quad
T_i= -\fft{r_i(1-r_i^2/l^2)}{2\pi} \sum_k \fft1{r_i^2+a_k^2} +
  \fft{1+r_i^2/l^2}{4\pi r_i}\,,\nonumber\\
\Omega_i^k &=& \fft{(1-r_i^2/l^2) a_k}{r_i^2 + a_k^2}\,,\quad
V_i = \frac{r_i A_i}{d-1}\left[1+\frac{1-r_i^2/l^2}{(d-2)r_i^2}\sum_k \frac{a_k^2}{\Xi_k}\right]\,.
\ea

%%%%%%%%%%%%%%%%%%%%%%%%%%%%%%%%%%%%%%%%%%%
\subsection{Odd dimensions}
In odd dimensions ($d=2N+1$) the mass, temperature, and entropies get modified as follows:
\be
M=\frac{m \cA_{D-2}}{4\pi \prod_j \Xi_j}\left(\sum_k \frac{1}{\Xi_k}-\frac{1}{2}\right)\,,
\ee
\ba\label{TDodd}
S_c&=& \fft{\cA_{d-2}}{4r_c}\, \prod_k \fft{r_c^2+a_k^2}{\Xi_k}=\frac{A_c}{4}\,,\quad
T_c= -\fft{r_c(1-r_c^2/l^2)}{2\pi} \sum_k \fft1{r_c^2+a_k^2} +\fft1{2\pi r_c}\,,\quad\\
S_b&=& \fft{\cA_{d-2}}{4r_b}\, \prod_k \fft{r_b^2+a_k^2}{\Xi_k}=\frac{A_b}{4}\,,\quad
T_b= \fft{r_b(1-r_b^2/l^2)}{2\pi} \sum_k \fft1{r_b^2+a_k^2} -\fft1{2\pi r_b}\,,\quad\\
S_i&=& \fft{\cA_{d-2}}{4r_i}\, \prod_k \fft{r_i^2+a_k^2}{\Xi_k}=\frac{A_i}{4}\,,\quad
T_i= -\fft{r_i(1-r_i^2/l^2)}{2\pi} \sum_k \fft1{r_i^2+a_k^2} +\fft1{2\pi r_i}\,.\quad
\ea
All other quantities ($\Omega$'s, $J$'s, and $V$'s) remain as in the even-dimensional case.
It is easy to verify that in both even and odd dimensions these quantities satisfy the first laws \eqref{firstBHb}--\eqref{firstBHi} as well as the Smarr formulae \eqref{Smarr1}--\eqref{Smarr2}.

The asymptotically AdS black holes are formally obtained by setting $l\to il$. The corresponding thermodynamics in extended phase space has been studied e.g. in \cite{Altamirano:2013ane, Altamirano:2013uqa, Altamirano:2014tva, Dolan:2014jva}. In particular it has been shown that for $d\leq 5$ one only observes the Van der Waals-like phase transition, similar to the spherically symmetric charged AdS case.
More interesting is the behavior of black holes in $d\geq 6$ dimensions. In particular, in $d=6$ dimensions one can observe the reentrant phase transitions \cite{Altamirano:2013ane}, triple points  \cite{Altamirano:2013uqa}, or Van der Waals behavior, dependent on the ratio of the two angular momenta $q=J_1/J_2$. Interestingly, as we have seen in the main text, the reentrant phase transition for $d=6$ dimensional black holes can also occur for the dS case.

\providecommand{\href}[2]{#2}\begingroup\raggedright
\endgroup


\begin{thebibliography}{10}

\bibitem{bekenstein1973black}
J.~D. Bekenstein, {\it Black holes and entropy},  {\em Physical Review D} {\bf
  7} (1973), no.~8 2333.

\bibitem{Bardeen:1973gs}
J.~M. Bardeen, B.~Carter, and S.~Hawking, {\it {The Four laws of black hole
  mechanics}},  {\em Commun.Math.Phys.} {\bf 31} (1973) 161--170.

\bibitem{hawking1983thermodynamics}
S.~W. Hawking and D.~N. Page, {\it Thermodynamics of black holes in anti-de
  sitter space},  {\em Communications in Mathematical Physics} {\bf 87} (1983),
  no.~4 577--588.




\bibitem{Chamblin:1999tk}
A.~Chamblin, R.~Emparan, C.~V. Johnson, and R.~C. Myers, {\it {Charged AdS
  black holes and catastrophic holography}},  {\em Phys.Rev.} {\bf D60} (1999)
  064018, [\href{http://arxiv.org/abs/hep-th/9902170}{{\tt hep-th/9902170}}].

\bibitem{Chamblin:1999hg}
A.~Chamblin, R.~Emparan, C.~V. Johnson, and R.~C. Myers, {\it {Holography,
  thermodynamics and fluctuations of charged AdS black holes}},  {\em
  Phys.Rev.} {\bf D60} (1999) 104026,
  [\href{http://arxiv.org/abs/hep-th/9904197}{{\tt hep-th/9904197}}].

\bibitem{Cvetic:1999ne}
M.~Cvetic and S.~S. Gubser, {\it {Phases of R charged black holes, spinning
  branes and strongly coupled gauge theories}},  {\em JHEP} {\bf 9904} (1999)
  024, [\href{http://arxiv.org/abs/hep-th/9902195}{{\tt hep-th/9902195}}].

\bibitem{Kubiznak:2012wp}
D.~Kubiznak and R.~B. Mann, {\it {P-V criticality of charged AdS black holes}},
   {\em JHEP} {\bf 1207} (2012) 033,
  [\href{http://arxiv.org/abs/1205.0559}{{\tt arXiv:1205.0559}}].

\bibitem{Altamirano:2013ane}
N.~Altamirano, D.~Kubiznak, and R.~B. Mann, {\it {Reentrant phase transitions
  in rotating anti–de Sitter black holes}},  {\em Phys.Rev.} {\bf D88}
  (2013), no.~10 101502, [\href{http://arxiv.org/abs/1306.5756}{{\tt
  arXiv:1306.5756}}].

\bibitem{Altamirano:2013uqa}
N.~Altamirano, D.~Kubiznak, R.~B. Mann, and Z.~Sherkatghanad, {\it {Kerr-AdS
  analogue of triple point and solid/liquid/gas phase transition}},  {\em
  Class.Quant.Grav.} {\bf 31} (2014) 042001,
  [\href{http://arxiv.org/abs/1308.2672}{{\tt arXiv:1308.2672}}].

\bibitem{Altamirano:2014tva}
N.~Altamirano, D.~Kubiznak, R.~B. Mann, and Z.~Sherkatghanad, {\it
  {Thermodynamics of rotating black holes and black rings: phase transitions
  and thermodynamic volume}},  {\em Galaxies} {\bf 2} (2014) 89--159,
  [\href{http://arxiv.org/abs/1401.2586}{{\tt arXiv:1401.2586}}].

\bibitem{Dolan:2014jva}
B.~P. Dolan, {\it {Black holes and Boyle's law — The thermodynamics of the
  cosmological constant}},  {\em Mod.Phys.Lett.} {\bf A30} (2015), no.~03n04
  1540002, [\href{http://arxiv.org/abs/1408.4023}{{\tt arXiv:1408.4023}}].

%\cite{Kubiznak:2016qmn}
\bibitem{Kubiznak:2016qmn}
  D.~Kubiznak, R.~B.~Mann and M.~Teo,
  {\it Black hole chemistry: thermodynamics with Lambda,}
  \href{http://arxiv.org/abs/1608.06147}{{\tt arXiv:1608.06147}}.
  %arXiv:1608.06147 [hep-th].
  %%CITATION = ARXIV:1608.06147;%%


\bibitem{Witten:1998zw}
E.~Witten, {\it {Anti-de Sitter space, thermal phase transition, and
  confinement in gauge theories}},  {\em Adv.Theor.Math.Phys.} {\bf 2} (1998)
  505--532, [\href{http://arxiv.org/abs/hep-th/9803131}{{\tt hep-th/9803131}}].

\bibitem{Strominger:2001pn}
A.~Strominger, {\it {The dS / CFT correspondence}},  {\em JHEP} {\bf 0110}
  (2001) 034, [\href{http://arxiv.org/abs/hep-th/0106113}{{\tt
  hep-th/0106113}}].

\bibitem{Ashtekar:2014zfa}
A.~Ashtekar, B.~Bonga, and A.~Kesavan, {\it {Asymptotics with a positive
  cosmological constant: I. Basic framework}},  {\em Class. Quant. Grav.} {\bf
  32} (2015), no.~2 025004, [\href{http://arxiv.org/abs/1409.3816}{{\tt
  arXiv:1409.3816}}].

\bibitem{Ashtekar:2015lla}
A.~Ashtekar, B.~Bonga, and A.~Kesavan, {\it {Asymptotics with a positive
  cosmological constant: II. Linear fields on de Sitter space-time}},
  \href{http://arxiv.org/abs/1506.06152}{{\tt arXiv:1506.06152}}.

\bibitem{Cai:2001sn}
R.-G. Cai, {\it {Cardy-Verlinde formula and asymptotically de Sitter spaces}},
  {\em Phys.Lett.} {\bf B525} (2002) 331--336,
  [\href{http://arxiv.org/abs/hep-th/0111093}{{\tt hep-th/0111093}}].

\bibitem{Cai:2001tv}
R.-G. Cai, {\it {Cardy-Verlinde formula and thermodynamics of black holes in de
  Sitter spaces}},  {\em Nucl.Phys.} {\bf B628} (2002) 375--386,
  [\href{http://arxiv.org/abs/hep-th/0112253}{{\tt hep-th/0112253}}].

\bibitem{Sekiwa:2006qj}
Y.~Sekiwa, {\it {Thermodynamics of de Sitter black holes: Thermal cosmological
  constant}},  {\em Phys.Rev.} {\bf D73} (2006) 084009,
  [\href{http://arxiv.org/abs/hep-th/0602269}{{\tt hep-th/0602269}}].

\bibitem{Dolan:2013ft}
B.~P. Dolan, D.~Kastor, D.~Kubiznak, R.~B. Mann, and J.~Traschen, {\it
  {Thermodynamic Volumes and Isoperimetric Inequalities for de Sitter Black
  Holes}},  {\em Phys.Rev.} {\bf D87} (2013), no.~10 104017,
  [\href{http://arxiv.org/abs/1301.5926}{{\tt arXiv:1301.5926}}].

\bibitem{Gomberoff:2003ea}
A.~Gomberoff and C.~Teitelboim, {\it {de Sitter black holes with either of the
  two horizons as a boundary}},  {\em Phys. Rev.} {\bf D67} (2003) 104024,
  [\href{http://arxiv.org/abs/hep-th/0302204}{{\tt hep-th/0302204}}].

\bibitem{Cvetic:2010mn}
M.~Cvetic, G.~W. Gibbons, and C.~N. Pope, {\it {Universal Area Product Formulae
  for Rotating and Charged Black Holes in Four and Higher Dimensions}},  {\em
  Phys. Rev. Lett.} {\bf 106} (2011) 121301,
  [\href{http://arxiv.org/abs/1011.0008}{{\tt arXiv:1011.0008}}].

\bibitem{Castro:2012av}
A.~Castro and M.~J. Rodriguez, {\it {Universal properties and the first law of
  black hole inner mechanics}},  {\em Phys. Rev.} {\bf D86} (2012) 024008,
  [\href{http://arxiv.org/abs/1204.1284}{{\tt arXiv:1204.1284}}].

\bibitem{Page:2015gia}
D.~N. Page and A.~A. Shoom, {\it {The Universal Area Product: An Heuristic
  Argument}},
  %\cite{Page:2015gia}
  {\em Phys.\ Rev.}  {\bf D92},  (2015) 044039,
  [\href{http://arxiv.org/abs/1504.05581}{{\tt arXiv:1504.05581}}].

\bibitem{Ghezelbash:2001vs}
A.~Ghezelbash and R.~B. Mann, {\it {Action, mass and entropy of
  Schwarzschild-de Sitter black holes and the de Sitter / CFT correspondence}},
   {\em JHEP} {\bf 0201} (2002) 005,
  [\href{http://arxiv.org/abs/hep-th/0111217}{{\tt hep-th/0111217}}].



\bibitem{Kastor:2009wy}
D.~Kastor, S.~Ray, and J.~Traschen, {\it {Enthalpy and the Mechanics of AdS
  Black Holes}},  {\em Class.Quant.Grav.} {\bf 26} (2009) 195011,
  [\href{http://arxiv.org/abs/0904.2765}{{\tt arXiv:0904.2765}}].



\bibitem{Urano:2009xn}
M.~Urano, A.~Tomimatsu, and H.~Saida, {\it {Mechanical First Law of Black Hole
  Spacetimes with Cosmological Constant and Its Application to Schwarzschild-de
  Sitter Spacetime}},  {\em Class.Quant.Grav.} {\bf 26} (2009) 105010,
  [\href{http://arxiv.org/abs/0903.4230}{{\tt arXiv:0903.4230}}].

\bibitem{Ma:2013aqa}
M.-S. Ma, H.-H. Zhao, L.-C. Zhang, and R.~Zhao, {\it {Existence condition and
  phase transition of Reissner-Nordström-de Sitter black hole}},  {\em
  Int.J.Mod.Phys.} {\bf A29} (2014) 1450050,
  [\href{http://arxiv.org/abs/1312.0731}{{\tt arXiv:1312.0731}}].

\bibitem{Zhao:2014zea}
R.~Zhao, M.~Ma, H.~Zhao, and L.~Zhang, {\it {The Critical Phenomena and
  Thermodynamics of the Reissner-Nordstrom-de Sitter Black Hole}},  {\em
  Adv.High Energy Phys.} {\bf 2014} (2014) 124854.

\bibitem{Zhao:2014raa}
H.-H. Zhao, L.-C. Zhang, M.-S. Ma, and R.~Zhao, {\it {P-V criticality of higher
  dimensional charged topological dilaton de Sitter black holes}},  {\em
  Phys.Rev.} {\bf D90} (2014), no.~6 064018.

\bibitem{Ma:2014hna}
M.-S. Ma, L.-C. Zhang, H.-H. Zhao, and R.~Zhao, {\it {Phase transition of the
  higher dimensional charged Gauss-Bonnet black hole in de Sitter spacetime}},
  {\em Adv.High Energy Phys.} {\bf 2015} (2015) 134815,
  [\href{http://arxiv.org/abs/1410.5950}{{\tt arXiv:1410.5950}}].

\bibitem{Guo:2015waa}
X.~Guo, H.~Li, L.~Zhang, and R.~Zhao, {\it {Thermodynamics and phase transition
  in the Kerr–de Sitter black hole}},  {\em Phys.Rev.} {\bf D91} (2015),
  no.~8 084009.


%\cite{Li:2016zca}
\bibitem{Li:2016zca}
  H.~F.~Li, M.~S.~Ma and Y.~Q.~Ma,
  {\it Thermodynamic properties of black holes in de Sitter space,}
  %Submitted to: PTEP (2015)
  \href{http://arxiv.org/abs/1605.08225}{{\tt arXiv:1605.08225}}.
  %[arXiv:1605.08225 [hep-th]].
  %%CITATION = ARXIV:1605.08225;%%
  %2 citations counted in INSPIRE as of 24 Aug 2016


\bibitem{McInerney:2015xwa}
  J.~McInerney, G.~Satishchandran and J.~Traschen,
  {\it Cosmography of KNdS Black Holes and Isentropic Phase Transitions},
  {\em Class.\ Quant.\ Grav.\ }  {\bf 33} (2016) no.10,  105007,
  %doi:10.1088/0264-9381/33/10/105007
  [\href{http://arxiv.org/abs/1509.02343}{{\tt arXiv:1509.02343}}].



\bibitem{Kastor:1992nn}
D.~Kastor and J.~H. Traschen, {\it {Cosmological multi - black hole
  solutions}},  {\em Phys.Rev.} {\bf D47} (1993) 5370--5375,
  [\href{http://arxiv.org/abs/hep-th/9212035}{{\tt hep-th/9212035}}].

\bibitem{Bhattacharya:2015mja}
S.~Bhattacharya, {\it {A note on entropy of de Sitter black holes}},
{\em Eur.\ Phys.\ J.}  {\bf C76} (2016) 112,
 [\href{http://arxiv.org/abs/1506.07809}{{\tt arXiv:1506.07809}}].

\bibitem{FilJa}
D.~Kubiznak, I.~Shehzad, and F.~Simovic, {\it {On effective temperature of de Sitter black
  holes}},  {\em in preparation} (2016).



\bibitem{ginsparg1983semiclassical}
P.~Ginsparg and M.~J. Perry, {\it Semiclassical perdurance of de sitter space},
   {\em Nuclear Physics B} {\bf 222} (1983), no.~2 245--268.



\bibitem{nariai1951new}
H.~Nariai, {\it On a new cosmological solution of einstein's fieldequations of
  gravitation},  {\em Science reports of the Tohoku University 1st ser.
  Physics, chemistry, astronomy} {\bf 35} (1951), no.~1 62--67.


\bibitem{Mann:1995vb}
R.~B. Mann and S.~F. Ross, {\it {Cosmological production of charged black hole
  pairs}},  {\em Phys.Rev.} {\bf D52} (1995) 2254--2265,
  [\href{http://arxiv.org/abs/gr-qc/9504015}{{\tt gr-qc/9504015}}].

\bibitem{Booth:1998gf}
I.~Booth and R.~B. Mann, {\it {Cosmological pair production of charged and
  rotating black holes}},  {\em Nucl.Phys.} {\bf B539} (1999) 267--306,
  [\href{http://arxiv.org/abs/gr-qc/9806056}{{\tt gr-qc/9806056}}].

\bibitem{Anninos:2010gh}
D.~Anninos and T.~Anous, {\it {A de Sitter Hoedown}},  {\em JHEP} {\bf 08}
  (2010) 131, [\href{http://arxiv.org/abs/1002.1717}{{\tt arXiv:1002.1717}}].

%\cite{Gibbons:2004uw}
\bibitem{Gibbons:2004uw}
  G.~W.~Gibbons, H.~Lu, D.~N.~Page and C.~N.~Pope,
  {\it The General Kerr-de Sitter metrics in all dimensions,}
  J.\ Geom.\ Phys.\  {\bf 53}, 49 (2005),
  [\href{http://arxiv.org/abs/hep-th/0404008}{{\tt hep-th/0404008}}].


%\cite{Gibbons:2004js}
\bibitem{Gibbons:2004js}
  G.~W.~Gibbons, H.~Lu, D.~N.~Page and C.~N.~Pope,
  {\it Rotating black holes in higher dimensions with a cosmological constant},
  Phys.\ Rev.\ Lett.\  {\bf 93} (2004) 171102,
  [\href{http://arxiv.org/abs/hep-th/0409155}{{\tt hep-th/0409155}}].


%\cite{Gibbons:2004ai}
\bibitem{Gibbons:2004ai}
  G.~W.~Gibbons, M.~J.~Perry and C.~N.~Pope,
  {\it The First law of thermodynamics for Kerr-anti-de Sitter black holes},
  Class.\ Quant.\ Grav.\  {\bf 22}, 1503 (2005)
  [\href{http://arxiv.org/abs/hep-th/0408217}{{\tt hep-th/0408217}}].
 %1088/0264-9381/22/9/002;%%
  %228 citations counted in INSPIRE as of 18 Jan 2016




\end{thebibliography}
\end{document}